\documentclass[journal]{IEEEtran}

\usepackage{courier}
\usepackage[usenames,dvipsnames]{color} 
\usepackage{color,url,xspace}
\usepackage{amssymb,amsmath}
\usepackage{graphicx}
\usepackage{colordvi}
\usepackage{epsfig}
\usepackage{endnotes}
\usepackage{verbatim}
\usepackage{subfigure}
\usepackage{textcomp}
\usepackage{subfig}
\usepackage{setspace}
\usepackage{xcolor}
\usepackage{amsmath}
\usepackage{cite}
\usepackage{times}
\usepackage{booktabs, multirow}
\usepackage{amsthm}
\usepackage{nccmath}
\usepackage{algorithmicx,algpseudocode}
\usepackage{algpseudocode,algorithm,algorithmicx}
\usepackage{caption}
\usepackage{mathtools}

\newcommand*\Let[2]{\State #1 $\gets$ #2}
\algrenewcommand\algorithmicrequire{\textbf{Precondition:}}
\algrenewcommand\algorithmicensure{\textbf{Postcondition:}}

\usepackage{color}
\usepackage{soul}

\begin{document}
	\title{Blind Signal Classification for Non-Orthogonal Multiple Access in Vehicular Networks}
	\author{
		\IEEEauthorblockN{
			Minseok Choi, \IEEEmembership{Member, IEEE,}
			Daejung Yoon, \IEEEmembership{Member, IEEE,}
			and
			Joongheon Kim, \IEEEmembership{Senior Member, IEEE}
		}
		\thanks{This research was supported by Institute for Information \& Communications Technology Promotion (IITP) grant funded by the Korea government (MSIT) (No.2018-0-00170, Virtual Presence in Moving Objects through 5G).}
		\thanks{M. Choi is with University of Southern California, Los Angeles, CA 90089, USA (e-mail: choimins@usc.edu).}
		\thanks{D. Yoon is with Nokia Bell Labs, Nozay 91620, France (e-mail: yoondanny@hotmail.com).}
		\thanks{J. Kim is with the School of Electrical Engineering, Korea University, Seoul 02841, South Korea (e-mail: joongheon@korea.ac.kr).}
		\thanks{J. Kim is the corresponding author of this paper.}
	}
	
	\markboth{IEEE Transactions on Vehicular Technology (Submission)}%
	{IEEE Transactions on Vehicular Technology (Submission)}
	
	\maketitle
	
	\begin{abstract}
		In this paper, blind signal classification and detection in a non-orthogonal multiple access (NOMA) system are explored.
		Since a NOMA scheme superposes the multiple-user (MU) signals within nonorthogonal resources, classical modulation classification methods used in orthogonal multiple access (OMA) systems are not sufficient to process the superposed NOMA signal.
		NOMA receivers require information about the multiple access schemes such as modulation order and need interference cancellation of the co-scheduled user's signal; therefore, a NOMA system causes more high-layer signaling overheads than OMA during packet scheduling. 
		Blind detection algorithms used for multiplexing information are considered to be possible solutions; however, they pose various challenges and could cause performance loss while performing blind modulation classification in the order of 1) OMA/NOMA classification, 2) co-scheduled user's modulation classification, and 3) classification of the signal, due to the necessity for successive interference cancellation (SIC). 
		To improve the performance of blind detection, we propose a NOMA transmission scheme that applies phase-rotation to data or pilot symbols depending on the NOMA multiplexing format, as an aid to the blind detection.  
		The proposed classification algorithm can implicitly provide essential information on NOMA multiplexing without the need for any extra high layer signaling or resources. 
		The performance improvement is verified through simulation studies, and it is found that the proposed algorithm provides a gain of more than 1 dB compared to the existing blind signal classification methods and shows almost equivalent performance as the genie information scheme. 
	\end{abstract}
	
	
	\IEEEpeerreviewmaketitle
	
	\section{Introduction}
	\label{sec:Introduction}
	
	To utilize the radio spectrum efficiently for a massive number of user terminals (UTs), non-orthogonal multiple access (NOMA) based on power multiplexing has been widely studied \cite{OMA_vs_NOMA:Wang,TVT:1,TVT:2,NOMA_basic:Docomo-Saito,NOMA_basic:Higuchi}.
	In particular, NOMA has been actively researched as a promising technology to improve system performance in 5G networks\cite{NOMA_5G:Dai,NOMA_5G:Ding,NOMA_5G:Islam} and to provide robustness in high-mobility vehicular networks~\cite{NOMA_VN:Di,NOMA_VN:Qian}.
	The 3rd Generation Partnership Project (3GPP) has studied deployment scenarios and receiver designs for NOMA systems in Rel-14 in the context of a working item labeled multiple user superposition transmission (MuST)\cite{MuST:3GPP}.
	
	NOMA superposes the multiple-user (MU) signals within the same frequency, time, or spatial domain, and therefore successive interference cancellation (SIC) is generally considered to detect non-orthogonally multiplexed signals \cite{NOMA_basic:Docomo-Saito,NOMA_basic:Higuchi}.
	Theoretically, NOMA is known to provide significant benefits in terms of improving the cell throughput\cite{Book:Tse}; nevertheless, such gains can be obtained only when the receiver is able to cancel or sufficiently suppress interference from co-scheduled users.
	NOMA has been extensively researched in conjunction with various technologies on the basis of the well-designed SIC. 
	Studies have been conducted on a system that applies NOMA to multiple-input multiple-output (MIMO) systems \cite{NOMA-MIMO:TWC2016Ding, NOMA-MIMO:TWC2016Ding2}, and studies on NOMA in cooperative networks were reported in \cite{NOMA:CL2015Ding, NOMA:TWC2018Choi}.
	In addition, a distributed NOMA scheme without the requirement of SIC for the Internet of Things (IoT) was studied in \cite{IOTJ2019Liu2}.
	In most existing studies on NOMA, an ideal SIC with a knowledge of the channel state information (CSI) was assumed, but in the recent studies in \cite{NOMA:TVT2018Gui,IOTJ2019Liu}, imperfect CSI for NOMA was handled using deep learning. 
	
	For orthogonal multiple access (OMA), there have been extensive research efforts on blind modulation classification (MC). 
	MC was originally developed for military applications such as electronic warfare; therefore, most of the existing MC techniques were developed for systems having no knowledge of the signal amplitude, phase, channel fading characteristics, and noise distribution\cite{MC-ML-unknown1:Hameed,MC-ML-unknown2:Erfan}.
	In \cite{MC-ML:Wei}, a maximum likelihood (ML)-based classifier was presented to provide optimal performance in the presence of white Gaussian noise when candidate modulation schemes are equally probable.
	However, ML-based classification requires high computational complexity; therefore, the feature-based approaches for blind MC presented in \cite{MC-feature:ICCSPA2013Hazza} take advantage of the fact that good statistical features allow robust blind MC.
	Furthermore, convolutional neural network (CNN)-based feature extraction methods were recently addressed in \cite{MC-ML:TCCN2018Rajendran, CNN-MC:DySPAN2017West, CNN-MC:TVT2018Meng, CNN-MC:TCCN2018Tian}.
	However, the existing studies on blind signal classification were primarily limited to MC techniques in an OMA system.
	
	As NOMA has become popular and is being implemented in many applications, blind MC is also important in the NOMA system. 
	Since the MU signals are superposed in the NOMA system, additional and necessary information of the signal modulation is required. 
	Before attempting to handle interference, the NOMA receiver must first determine the presence of co-scheduled users. 
	If SIC is to be used, the modulation order and power allocation ratio of the co-scheduled users should also be known to the receiver. 
	In summary, the blind MC steps required for the NOMA receiver are as follows: 1) OMA/NOMA classification, 2) co-scheduled UTs' MC, and 3) classification of the signal based on the necessity of SIC.
	Therefore, we refer to these classification steps in the NOMA system as blind signal classification, not to be confused with blind MC for OMA. 
	Blind classification of something else, e.g., channel coding rate, can be considered, but this paper basically focuses on blind MC in the NOMA system; therefore, the classification steps for information not related to signal modulation are not considered in this paper.
	
	The information required for signal classification and data detection can be transmitted to the receiver via a high layer; however, the required signaling overhead is a concern as more information of the signal is needed for decoding at the NOMA receiver. 
	In particular, in vehicular networks with limited energy and resources (e.g., time limits due to the high mobility in vehicular networks) \cite{NOMA_VN:Di,NOMA_VN:Qian}, this motivates the use of blind signal classification at the receiver side followed by appropriate data detection.
	Moreover, vehicular networks must cope with periodic short burst communications related to safety information and alarm services \cite{CommMag2013Araniti, TVT2016Kim}. 
	Because the concern about signaling overheads becomes more critical in the case of short burst communications, blind signal classification could emerge as a promising technology to reduce these 
	overheads in vehicular communications \cite{TVT2017Majhi}.
	
	While the existing MC techniques are aimed at determining the UT's modulation, the NOMA receiver attempts to classify the co-scheduled user's modulation to perform SIC.
	In addition, the NOMA system should perform OMA/NOMA classification, as well as determine whether SIC is required for the received signal.
	The blind MC technique in the context of NOMA has been actively studied in 3GPP, for example, \cite{MuST:3GPP} provides throughput analysis of ML-based blind MC in NOMA systems.
	However, the improved blind MC technique has not been well explored in academia. 
	The recent study in \cite{ArXiv2019Choi} was focused on signal classification only with respect to the necessity of SIC, and both power allocation and the user scheduling scheme, which guarantee a reliable classification performance, were jointly optimized.
	In this paper, the performance effect of errors in blind NOMA signal classification is analyzed and the ensuing receiver challenges in practical MU cases are addressed.
	In addition, two transmission policies are proposed for improving the performances of ML-based blind NOMA signal classification.
	
	The main contributions of this paper are summarized as follows:
	\begin{itemize}
		
		\item Signal to interference plus noise ratio (SINR) and user capacity analyses to determine the effect of errors on performance in the three blind NOMA signal classification steps are presented.
		
		\item A phase-rotated modulation is proposed for blind NOMA signal classification.
		The rotated data symbols implicitly render the constellations of modulation formats as blind detection aids.
		This method is based on the existing ML-based classification algorithm \cite{MC-ML:TCCN2018Rajendran}.
		
		\item A pilot-rotation transmission method and the corresponding new signal classification algorithm are proposed.
		By using the algorithm, a receiver estimates the rotation value of pilots and utilizes it for blind signal classification.
		Since the proposed scheme depends only on the rotated phases of the pilots and not on the pilot values, it requires no extra pilot overhead. 
		
		\item The presented numerical results verify the performance analysis of blind signal classification in the NOMA system.
		Moreover, the proposed phase-rotated modulation and the pilot-rotation transmission scheme are shown to provide better classification performances than the conventional ML-based method.
		
	\end{itemize}
	
	The rest of the paper is organized as follows. 
	The NOMA system model and the blind classification steps for received NOMA signals are described in Section \ref{sec:Preliminaries}.
	SINR analysis for the three steps of blind NOMA signal classification and the capacity of a NOMA UT in  the presence of signal classification errors are provided in Sections \ref{sec:SINR_analysis} and \ref{sec:Capacity_analysis_for_blind_modulation_Detection}, respectively.
	The proposed phase-rotated modulation is described in Section \ref{sec:phase_rotated_modulation}.
	In Section \ref{sec:Pilot-Reuse_based_blind_OMA/NOMA_and_Modualtion_detection}, the proposed pilot-rotation transmission method and the corresponding new signal classification algorithm are presented.
	In Section \ref{sec:SimulationResults}, the performance improvements of the proposed algorithms are verified using the results of rigorous numerical simulations.
	
	\section{System Models}
	\label{sec:Preliminaries}
	
	\subsection{Non-Orthogonal Multiple Access Signal Model and Receiver Structure}
	
	In downlink power-multiplexing NOMA, a base station (BS) intentionally superposes the signals for multiple UTs with different power weightings.
	However, when information about the transmitted NOMA signal is unknown at the receiver side, the computational complexity of the ML-based signal classification in the NOMA system grows significantly with the number of co-scheduled UTs \cite{ArXiv2019Choi}.
	Therefore, this paper considers a two-user NOMA system. 
	The received signal in a two-user downlink NOMA transmission is given by
	\begin{equation}
	y = h(s_f + s_n) + w,
	\label{eq:received}
	\end{equation}
	where $y, s, h,$ and $w$ correspond to the received signal, transmitted symbol, channel gain, and thermal noise, respectively, and the subscripts $f$ and $n$ denote far and near UTs.
	In addition, $\mathbb{E}[|s_f|^2] = P_f$ and $\mathbb{E}[|s_n|^2] = P_n$, where $P_f$ and $P_n$ are the power allocations to two UTs.
	A BS normally schedules UTs having a large channel gain difference and allocates larger power to a far (weak) UT to compensate its low channel gain, i.e., $P_f>P_n$.
	Suppose that there is a normalized power constraint, $P_f + P_n = 1$.
	When its power allocation is large, the far UT does not perform SIC and only detects its data while ignoring the near UT's signal.
	Meanwhile, the near UT requires SIC to cancel the far UT's signal; therefore, only the near UT is considered as a NOMA-serviced user in general.
	For this reason, all the statements in this paper are focused on the near UT of the NOMA system.
	
	When the near UT performs SIC, interference, i.e., the far UT's signal, is regenerated from the decoder or the detector, corresponding to the codeword-level interference cancellation (CWIC) or symbol-level interference cancellation (SLIC), respectively \cite{VTC2015Yan}.
	In this study, CWIC is mainly utilized to mitigate the intra-cell interference unless otherwise noted; signal classification is required before CWIC is applied.
	
	Let $\mathcal{M}=\{M_0, M_{1}, \cdots, M_{L} \}$ be a set of modulation modes, including $L$ NOMA modes, $M_{l}$ for $l=1,\cdots,L$, and an OMA mode, $M_0$.
	The constellation set of the modulation mode $M_l$ is denoted by $\chi_l$ for all $l\in\{0,\cdots,L\}$. 
	For $l\in \{1,\cdots,L\}$, $\chi_l$ is constructed by combinations of power-scaled near and far UTs' constellation sets, $\chi_l = \chi_l^f \bigoplus \chi_l^n$, where $\chi_l^f$ and $\chi_l^n$ are power-scaled constellation sets of the near and far UTs, respectively.
	Therefore, the average powers of symbols in $\chi_l^f$ and $\chi_l^n$ are $P_f$ and $P_n$, respectively.
	In addition, let $\mathcal{N}$ be a set of the constellation points of all NOMA modes; 
	i.e., $\mathcal{N} = \chi_1 \cup \cdots \cup \chi_L$.
	
	\subsection{Maximum Likelihood-based Signal Classification}
	
	The existing ML-based MC algorithm \cite{MC-ML:Wei}, which is optimal in OMA based on hypotheses testing, can be directly applied to NOMA signal classification.
	We define some hypotheses to identify the received signal information:
	\begin{itemize}
		\item $\mathcal{H}_l$: the hypothesis of the signal modulated by the $l$-th mode $M_l$ for all $l \in \{0,1,\cdots,L\}$
		\item $\mathcal{H}_N$: the hypothesis of the signal, i.e., $\mathcal{H}_N = \mathcal{H}_1 \cup \cdots \cup \mathcal{H}_L$.
		\item $\mathcal{H}^f$: the hypothesis of the signal that does not require SIC.
		\item $\mathcal{H}^n$: the hypothesis of the signal for which SIC is necessary.
		\item $\mathcal{H}_{l}^f$: the hypothesis of the signal that is modulated by the $l$-th NOMA mode and does not require SIC for all $l\in \{ 1,\cdots,L\}$.
		\item $\mathcal{H}_{l}^n$: the hypothesis of the signal that is modulated by the $l$-th NOMA mode and requires SIC for all $l \in \{1,\cdots,L\}$.
	\end{itemize}
	
	The ML-based hypothesis testing can classify the received signal according to whether it is modulated by OMA or NOMA, which modulation and power weightings are used, and whether SIC is required or not.
	For example, suppose that the transmitted signal is modulated by the $l$-th modulation mode for $l\in\{1,\cdots,L\}$ and that the target signal can be decoded after performing SIC, i.e., $\mathcal{H}_l^n$ is true. 
	Then, the likelihood probability of hypothesis $\mathcal{H}_l^n$ is computed by
	\begin{equation}
	p(y| \mathcal{H}_l^n) = \frac{1}{|\chi_l|} \sum_{s \in \chi_l}{\frac{1}{\pi \sigma^2}e^{-\frac{| y - hs|^2}{\sigma^2}}},
	\end{equation}
	where $\sigma_n^2$ is the noise variance and $|\chi_l^n|$ is the number of symbols in $\chi_l^n$.
	If $K$ symbols are used for blind signal classification and are not correlated, the joint likelihood function of the $K$ symbols of $\mathbf{y}=[y_1, \cdots, y_K]$ is given by
	\begin{equation}
	\Gamma(\mathbf{y}|\mathcal{H}_l^n) = \prod_{k=1}^K p(y_k| \mathcal{H}_l^n).
	\end{equation}
	According to the ML criterion, the detected hypothesis $\hat{\mathcal{H}}$ can be determined by
	\begin{equation}
	\hat{\mathcal{H}} = \underset{\xi \in \mathcal{H}}{\arg\max}~ \Gamma( \mathbf{y} | \xi),
	\label{eq:ML-MC_whole}
	\end{equation}
	where $\mathcal{H}=\{ \mathcal{H}_0, \mathcal{H}_{1}^f, \cdots, \mathcal{H}_{L}^f, \mathcal{H}_{1}^n, \cdots, \mathcal{H}_{L}^n \}$.
	
	If $\hat{\mathcal{H}} = \mathcal{H}_0$, the receiver determines that the signal is modulated by OMA. 
	However, if $\hat{\mathcal{H}} = \mathcal{H}_{l}^f$, then the received signal is classified as a NOMA signal modulated by $M_l$, which does not require SIC.
	In addition, $\hat{\mathcal{H}} = \mathcal{H}_{l}^n$ represents that the received signal is a NOMA signal modulated by $M_l$ for which SIC is necessary.
	However, the accuracy of hypothesis testing is significantly degraded as the number of hypotheses grows. 
	Therefore, this paper considers the three-step classification framework for the NOMA signal to reduce the number of hypotheses in each classification step as follows: OMA/NOMA classification, MC (i.e., modulation order and power weightings), and near/far UT classification (i.e., the necessity for SIC).
	The relevant likelihood probabilities and the hypothesis testing results can be computed by 
	\subsubsection{OMA/NOMA Classification}
	
	\begin{figure} [t!]
		\centering
		\includegraphics[width=0.45\textwidth]{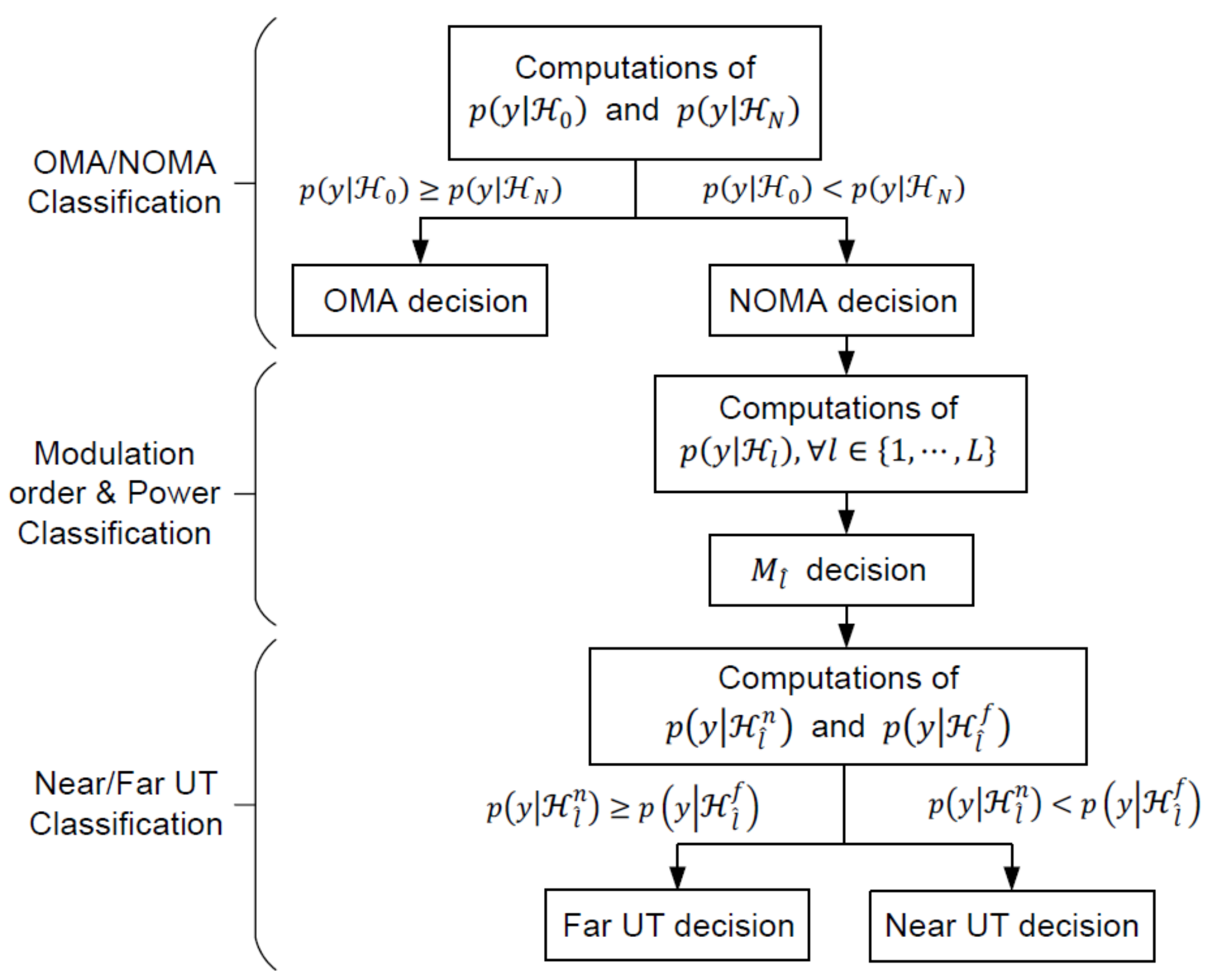}
		\caption{Processes of maximum likelihood-based signal classification in non-orthogonal multiple access systems}
		\label{Fig:MLPRM_scenario}
	\end{figure}
	
	\begin{align}
	p(y|\mathcal{H}_0) &= \frac{1}{|\chi_0|} \sum_{s \in \chi_0} {\frac{1}{\pi \sigma^2}e^{-\frac{|y-hs|^2}{\sigma^2}}}
	\label{MLprobOMA}\\
	p(y|\mathcal{H}_{N}) &= \frac{1}{|\mathcal{N}|} \sum_{s \in \mathcal{N}}{ \frac{1}{\pi \sigma^2}e^{-\frac{|y - hs|^2}{\sigma^2}} }
	\label{MLprobNOMA}\\
	\hat{\mathcal{H}} &= \underset{\xi \in \{\mathcal{H}_0, \mathcal{H}_N \}}{\arg\max} \Gamma (y|\xi);
	\label{ML OMA/NOMA}
	\end{align}
	\subsubsection{Modulation Classification}
	\begin{align}
	p(y|\mathcal{H}_{l}) &= \frac{1}{|\chi_l|} \sum_{s \in \chi_l}{ \frac{1}{\pi \sigma^2}e^{-\frac{|y - hs|^2}{\sigma^2}} }
	\label{MLProb:SumExp}\\
	\hat{\mathcal{H}} &= \underset{\xi \in\{ \mathcal{H}_1,\cdots,\mathcal{H}_L\}}{\arg\max} \Gamma (y|\xi);
	\label{MLdetectedMod:SumExp}
	\end{align}
	\subsubsection{Near/Far UT Classification}
	\begin{align}
	p(y|\mathcal{H}_{l}^n) &= \frac{1}{|\chi_{l}|} \sum_{s \in \chi_{l}}{\frac{1}{\pi \sigma^2}e^{-\frac{|y - hs|^2}{\sigma^2}}}
	\label{ML_nearUT}\\
	p(y|\mathcal{H}_{l}^f) &= \frac{1}{|\chi_{\hat{l}}^f|} \sum_{s \in \chi_{\hat{l}}^f}{ \frac{1}{\pi \sigma^2}e^{-\frac{|y - hs|^2}{\sigma^2}} } \\
	\hat{\mathcal{H}} &= \underset{\xi \in\{\mathcal{H}_l^n, \mathcal{H}_l^f \}}{\arg\max} \Gamma (y|\xi).
	\label{ML_farUT}
	\end{align}
	
	The overall steps of the ML-based signal classification in the NOMA system are shown in Fig. \ref{Fig:MLPRM_scenario}.
	In summary, OMA/NOMA classification should be performed first, 
	followed by the classification of the modulation orders and power ratios of the UTs.
	Near/far UT classification is the last step because it requires $\chi_l$ and $\chi_l^f$, whose modulation mode is already given.
	This paper investigates the additional classification steps required for NOMA compared to OMA; therefore, we assume that a UT already knows its own modulation order, whose classification was extensively studied earlier.
	Then, if the far and near UTs' modulation orders are different, the UT does not have to perform near/far UT classification;
	otherwise, near/far UT classification is necessary.
	
	The hierarchical classification steps can reduce the computational dimension and increase the accuracy of hypothesis testing as compared to that performed with respect to the whole set of modulation formats in \eqref{eq:ML-MC_whole}, in which $2L+1$ hypotheses are compared.
	In the proposed three-step classification framework, both OMA/NOMA and near/far UT classification compare only two hypotheses each, and $L$ hypotheses are compared in the MC step.
	In addition, a decrease in the number of compared hypotheses also reduces computational complexity.
	
	\section{SINR Analysis for Non-Orthogonal Multiple Access UT with Signal Classification Errors}
	\label{sec:SINR_analysis}
	
	In this section, the effects of signal classification errors on the SINR are examined.
	Again, the near UT is considered only as the NOMA-serviced user. 
	
	\subsection{OMA/NOMA and Near/Far UT Classification Errors}
	
	When the BS transmits a NOMA signal, but the near UT incorrectly classifies it as an OMA signal, severe performance degradation is expected.
	Even though the transmitted signal contains the far UT's signal component, incorrect OMA/NOMA classification causes the receiver to take no action to remedy the interference; i.e., it would not perform SIC.
	Similar results occur when an OMA signal is transmitted but the receiver classifies the signal as NOMA.
	In this case, the receiver performs SIC, but there is no interference in the OMA signal.
	Both cases do not guarantee a reliable classification performance. 
	Therefore, in this paper, the throughput is reasonably considered to be zero when an OMA/NOMA classification error occurs.
	
	Similarly, incorrect near/far UT classification significantly degrades the system performance.
	If incorrect near/far UT classification occurs, the far UT of the NOMA system cancels the target signal, and the near UT does not perform SIC.
	Therefore, an error in the near/far UT classification step is also assumed to yield no throughput. 
	The classification results of the far UT's modulation order and power ratio become meaningful only when the signal is classified as NOMA and near UT.
	
	\subsection{Power Ratio Classification Errors}
	\label{subsec:SINR_deg_PowerDetection}
	
	For the NOMA system, there are some modulation modes that have the same modulation orders but different power ratios for two UTs.
	The MC among these modes can be interpreted as power ratio classification.
	Although the receiver incorrectly classifies the power ratio as one of the competing modulation modes, the transmitted symbols can still be detected correctly if the incorrectly classified modulation mode has a constellation point indicating the same bit-labeling as the transmitted one.
	However, the SINR could be degraded because of erroneous power ratio classification.
	
	For simplicity, consider a flat fading channel and two competing modulation modes, $M_1$ and $M_2$, having the same order but different power allocation ratios for two NOMA UTs.
	Suppose that the transmitted signal is modulated by $M_1$; then, the received signal is given by
	\begin{equation}
	y = h( s_{f,1}(i) + s_{n,1}(k)) + w,
	\label{eq:received_nearUT}
	\end{equation}
	where $s_{f,1}(i)$ and $s_{n,1}(k)$ are the $i$-th and $k$-th symbols in $\chi_1^f$ and $\chi_1^n$ for the far and near UTs, respectively, and $\mathbb{E}_i[|s_{f,1}(i)|^2]=P_{f,1}$ and $\mathbb{E}_k[|s_{n,1}(k)|^2]=P_{n,1}$.
	
	Assuming perfect SIC, correct MC yields an SINR of
	\begin{equation}
	\eta_{1\rightarrow1} = \frac{P_{n,1}}{\tilde{\sigma}^2},
	\label{eq:SINR_n,1->1}
	\end{equation}
	where $\tilde{\sigma}^2 = \frac{\sigma^2}{|h|^2}$.
	The subscript $l\rightarrow m$ means that the transmitted mode is $M_l$, but $M_m$ is determined.
	Let $\hat{s}_f$ be the interference component for the near UT to be subtracted from the received signal \eqref{eq:received_nearUT}, which is regenerated by SIC.
	Then, the signal after SIC is denoted by
	\begin{align}
	y_{SIC} &= h(s_{f,1}(i)+ s_{n,1}(k) - \hat{s}_f) + w \\
	&= h s_{n,2}(k) + h (s_{f,1}(i)- \hat{s}_f) \nonumber \\
	&~~~+ h (s_{n,1}(k)-s_{n,2}(k)) + w.
	\label{eq:y_SIC}
	\end{align}
	Although the classification of power ratio is incorrect, the data detection could still be correct if $s_{n,2}(k)$, whose bit-labeling is the same as that of the transmitted signal $s_{n,1}(k)$, is detected.
	Therefore, the SINR becomes
	\begin{align}
	&\eta_{1\rightarrow2} = \nonumber \\ &~~~\frac{P_{n,2}}{\mathbb{E}_i[|s_{f,1}(i)-\hat{s}_f|^2]+\mathbb{E}_k[|s_{n,1}(k)-s_{n,2}(k)|^2]+\tilde{\sigma}^2}.
	\label{eq:SINR_n,1->2}
	\end{align}
	
	Note that, when $P_{n,1} \geq P_{n,2}$, incorrect classification of power ratio obviously results in SINR degradation; i.e., $\eta_{1\rightarrow 1} - \eta_{1\rightarrow 2} \geq 0$.
	However, if $P_{n,1} < P_{n,2}$, $\eta_{1\rightarrow 1} - \eta_{1\rightarrow 2} \geq 0$ only when
	\begin{align}
	P_{n,1}&(\mathbb{E}_i[|s_{f,1}(i)-\hat{s}_f|^2] + \mathbb{E}_k[|s_{n,1}(k)-s_{n,2}(k)|^2] + \tilde{\sigma}^2) \nonumber \\
	& \geq P_{n,2}\tilde{\sigma}^2.
	\label{eq:PowerDetection_condition}
	\end{align}
	If (\ref{eq:PowerDetection_condition}) is not satisfied, the SINR can increase even when the power ratio is incorrectly classified.
	Accordingly, power ratio classification becomes more important as the power allocated to the near UT, i.e., $P_{n,l}$, increases.
	
	Suppose that the same index $i$ of $s_{f,1}(i)$ and $s_{f,2}(i)$ indicates the same bit-labeling. 
	Then, when the power ratio classification is incorrect, it is highly likely that the interference regenerated by SIC is $\hat{s}_f = s_{f,2}(i)$ without a large $\tilde{\sigma}^2$.
	In this case, the inequality \eqref{eq:PowerDetection_condition} holds in the high SNR region.
	
	\subsection{Modulation Order Classification Errors}
	
	Consider two competing modes, $M_1$ and $M_2$, whose far UTs have different modulation orders. 
	It is very important for the near UT to find the modulation order of the far UT in order to perform SIC appropriately.
	If the modulation order of the far UT is incorrectly classified, the interference estimated by SIC, i.e., $\hat{s}_f$, also becomes incorrect.
	Then, the signal after SIC and the SINR of $\eta_{1 \rightarrow 2}$ are the same as \eqref{eq:y_SIC} and \eqref{eq:SINR_n,1->2}, respectively.
	
	Comparing with section \ref{subsec:SINR_deg_PowerDetection}, the only difference is that $\hat{s}_f$ is regenerated by SIC based on a competing mode whose modulation order is different from that of the transmitted mode, rather than power coefficient.
	In section \ref{subsec:SINR_deg_PowerDetection}, $\hat{s}_f$ can have the same bit-labeling as the transmitted symbol; however, $\hat{s}_f$ cannot because the modulation order of $M_2$ is different from that of the transmitted mode of $M_1$.
	
	Similar to the power ratio classification error described in Section \ref{subsec:SINR_deg_PowerDetection}, a classification error of the far UT's modulation order always causes SINR degradation at the near UT when $P_{n,1} \geq P_{n,2}$.
	However, when $P_{n,1} < P_{n,2}$, $\eta_{1\rightarrow 1}-\eta_{1\rightarrow 2} > 0$ only when the system satisfies the condition given by \eqref{eq:PowerDetection_condition}.
	Again, (\ref{eq:PowerDetection_condition}) can be satisfied in the high SNR region.
	If \eqref{eq:PowerDetection_condition} is not satisfied, the SINR is not degraded even when the far UT's modulation order is incorrectly classified. 
	However, this situation is considerably less likely to happen than when the power ratio is incorrectly classified but the SINR does not decrease.
	The reason is that $|s_{f,l}(i)-s_{f,u}(k)| > |s_{f,l}(i)-s_{f,v}(i)|$ in general, with the assumption that the far UT's modulation orders of $M_l$ and $M_u$ are different, while its modulation orders of $M_l$ and $M_v$ are the same but with different power ratios.
	
	An additional problem results from incorrect classification of a far UT's modulation order.
	As mentioned in Section \ref{sec:Preliminaries}, the decision feedback for SIC can be generated at either the symbol or codeword level.
	However, incorrect classification of the far UT's order does not allow the use of CWIC because of a mismatch in the codeword length.
	Because CWIC outperforms SLIC substantially, this is highly undesirable for the system-level performance.
	
	\section{Capacity of Non-Orthogonal Multiple Access UT with Signal Classification Errors}
	\label{sec:Capacity_analysis_for_blind_modulation_Detection}
	
	By performing SINR analysis, we can compute the capacity of a NOMA UT, which is denoted by $C$, including the effects of signal classification errors.
	With the assumption that the transmitted mode is $M_{l}$, let $p_{l \rightarrow m}$ be the probability that the classified modulation mode is $M_m$.
	The user capacity can then be computed as
	\begin{align}
	C = \sum\limits_{l=0}^L  \pi_l \mathbb{E}_h \Big[&\Big\{ p_{l \rightarrow l} q_l^n \log_2 (1+\eta_{l \rightarrow l}) \nonumber \\
	&~ + \sum_{m \neq l} p_{l \rightarrow m} q_m^n \log_2(1+\eta_{l \rightarrow m}) \Big\} \Big],
	\label{eq:capacity}
	\end{align}
	where $\pi_l$ is the probability that the signal is modulated by $M_l$ for all $l \in \{0,\cdots,L \}$ and $q_l^n$ is the probability that the signal of the modulation mode $M_l$ is determined to perform SIC.
	The equally probable modulation mode is assumed, i.e., $\pi_l=\frac{1}{L}$.
	Again, because only the near UT of the two-user NOMA system represents the NOMA user, the capacity in \eqref{eq:capacity} is achieved for the near UT case.
	In addition, $\eta_{l \rightarrow m}$ is achieved, as explained in Section \ref{sec:SINR_analysis}.
	Because incorrect OMA/NOMA and near/far classifications give rise to zero throughput, $\eta_{0\rightarrow l} = 0$, $\eta_{l\rightarrow 0} = 0$ for $l\neq0$, and only $q_l^n$ are considered, while the situation in which SIC is not necessary is not.
	
	The performance of ML-based signal classification strongly depends on how well the constellations of the competing modulation modes can be distinguished from one another.
	To quantify this effect, we denote the minimum distance between constellation sets of two different modulation modes of $M_l$ and $M_m$ by $d_{\text{min}}(M_l, M_m)$, $l \neq m$.
	In general, $d_{\text{min}}$ can be defined for $L$ modulation modes as follows:
	\begin{align}
	\forall s_1 \in \chi_1,~&\cdots,~\forall s_L \in \chi_L, \nonumber \\
	&d_{\text{min}}(M_1, ..., M_L) = \min d(s_1,...,s_L).
	\end{align}
	
	\begin{figure} [t!]
		\centering
		\includegraphics[width=0.3\textwidth]{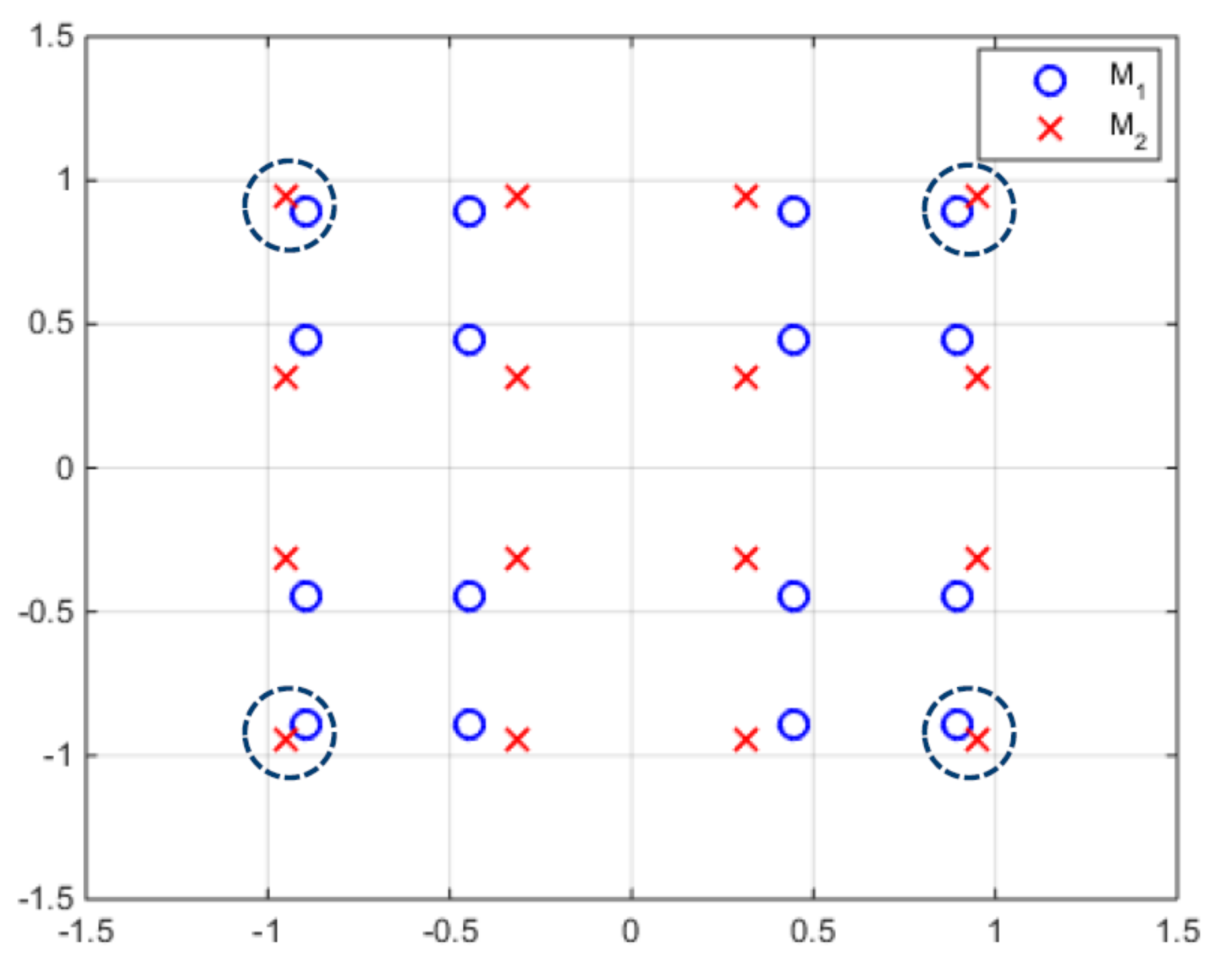}
		\caption{Legacy constellations of two modulation modes}
		\label{Fig:legacy_constellation}
	\end{figure}
	\begin{figure} [t!]
		\centering
		\includegraphics[width=0.3\textwidth]{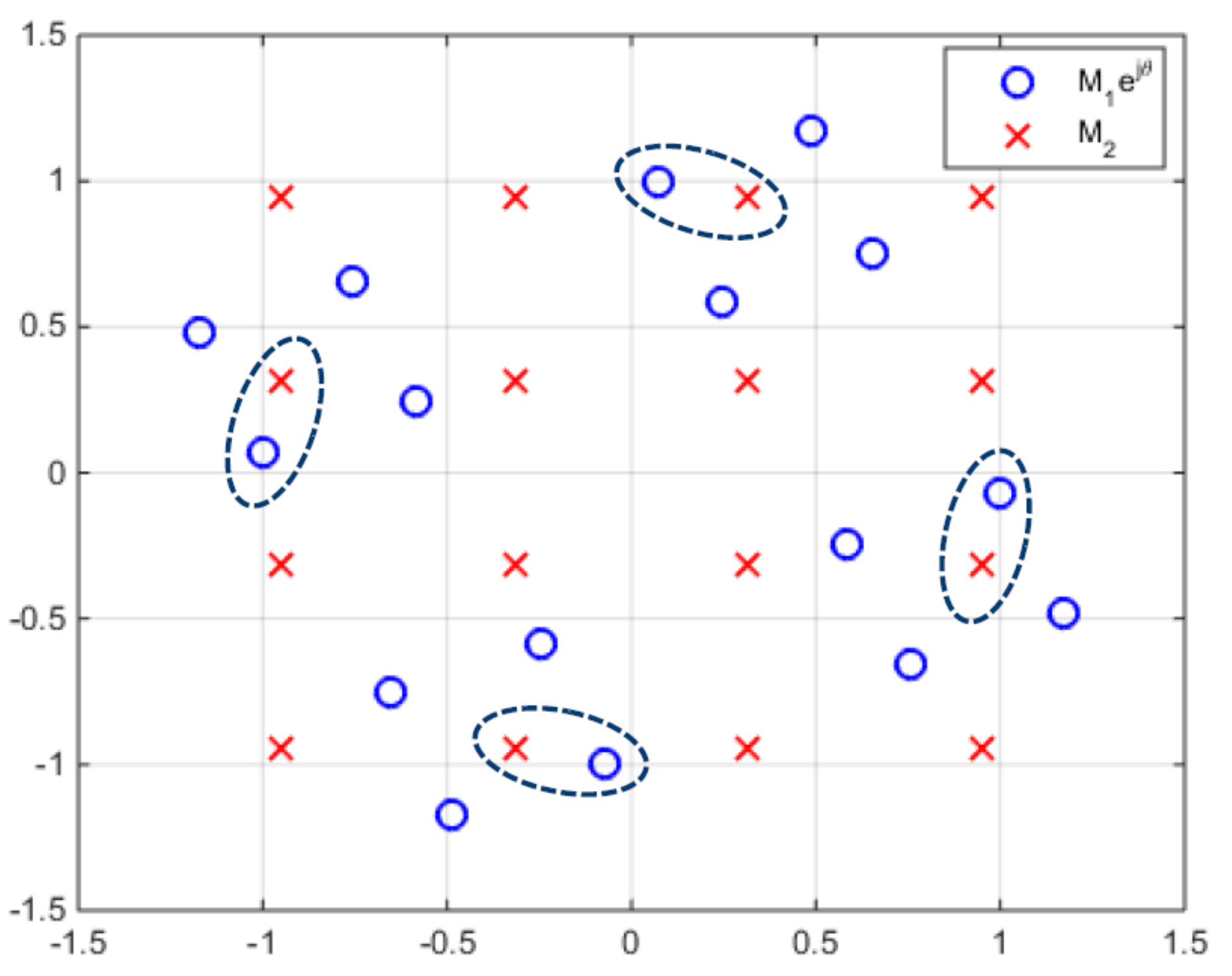}
		\caption{Rotated constellations of two modulation modes}
		\label{Fig:rotated_constellation}
	\end{figure}
	
	Fig. \ref{Fig:legacy_constellation} shows the constellation sets of two competing modulation modes, $M_1$ and $M_2$. 
	In this figure, $d_{\text{min}}(M_l, M_m)$ is the distance between two closest points from different modes, as marked by the dashed circles.
	These symbols are very close to each other; therefore, when ML-based signal classification is performed, these pairs are expected to be the main cause of incorrect classification.
	In this example, if all the symbol points are equally probable, the probability of a classification error can be computed as
	\begin{align}
	p_{l\rightarrow m} &= \frac{1}{|\chi_l| |\chi_m| } \sum_{i=1}^{|\chi_l|} \sum_{k=1}^{|\chi_m|} Q\Big(\frac{|h(s_l(i)-s_m(k))|/2}{\sigma/\sqrt{2}}\Big)
	\label{eq:modErrProb:allsymbol} \\
	&\approx \frac{N_{\text{min}}}{|\chi_l| |\chi_m|} Q\Big(\frac{h\cdot d_{\text{min}}(M_l, M_m)/2}{\sigma/\sqrt{2}}\Big),
	\label{eq:modErrProb:d_min}
	\end{align}
	where $s_l(i)$ is the $i$-th constellation point of $M_l$.
	
	In (\ref{eq:modErrProb:allsymbol}), let $s_m(k)$ be the closest to $s_l(i)$ among the constellation points of $M_k$. 
	Then, the case where $s_l(i)$ is confused with $s_m(k)$ becomes one of the most frequent classification errors.
	In this context, (\ref{eq:modErrProb:d_min}) is approximated one step further by considering only the symbol pairs from different modes, providing the minimum distance $d_{\text{min}}(M_l,M_m)$.
	In addition, $N_{\text{min}}$ is the total number of symbol pairs providing the minimum distance $d_{\text{min}}(M_l, M_m)$, and $N_{\text{min}}=4$ in Fig. \ref{Fig:legacy_constellation}.
	
	We remark some conclusions of analysis here. First, we can determine user capacity analytically by substituting the
	expressions for the SINR and classification error probability in (\ref{eq:capacity}). 
	Second, the trade-off between the classification error probability of $p_{l\rightarrow m}$ and the SINR degradation term of $\eta_{l\rightarrow m}$ is observed in \eqref{eq:capacity}. 
	According to \eqref{eq:modErrProb:d_min}, the classification error rate decreases with $d_{\text{min}}(M_l, M_m)$. 
	However, if an MC error occurs, further demodulation process got meaningless, as the SINR is dramatically degraded with increasing $d_{\text{min}}(M_l, M_m)$.
	Third, selection of the appropriate power allocation ratios for each NOMA modulation mode which maximizes capacity in \eqref{eq:capacity}, can be beneficial. 
	A maximum capacity is achieved with maximum power allocation on a near UT, but a far UT needs minimum power constrain in a reasonable problem setup.
	
	\section{Phase-Rotated Modulation\\ based on ML Signal Classification}
	\label{sec:phase_rotated_modulation}
	
	In this section, we propose a phase-rotated modulation method to increase the accuracy of signal classification.
	As explained in Section \ref{sec:Capacity_analysis_for_blind_modulation_Detection}, the reliability of ML-based signal classification strongly depends on how well the competing modes are distinguished from one another.
	Based on this observation, different phase rotations are assigned to individual modulation modes so that their constellation points are more effectively separated.
	
	A comparison of Fig. \ref{Fig:legacy_constellation} and Fig. \ref{Fig:rotated_constellation}, which show the legacy and phase-rotated composite constellations of two different modulation modes, respectively, demonstrates this idea.
	In Fig. \ref{Fig:rotated_constellation}, $M_1$ is rotated by $\theta$; therefore, it becomes $M_1e^{j\theta}$.
	The symbol pairs from different modes giving the minimum distance of $d_{\text{min}}(M_1e^{j\theta}, M_2)$ are marked by the dashed ellipses in Fig. \ref{Fig:rotated_constellation}.
	Because $d_{\text{min}}(M_1e^{j\theta}, M_2) > d_{\text{min}}(M_1, M_2)$, it is easily expected that the phase-rotated modulation provides a lower classification error probability than the legacy modulation method.
	If the same phase rotation is applied to every modulation mode, the minimum distance $d_{\text{min}}(M_1, M_2)$ between the rotated composite constellations of $M_1 e^{j\theta}$ and $M_2 e^{j\theta}$ remains unchanged.
	Accordingly, the application of different phase rotations to different modulation modes is a key point.
	We can make the phase list $\Theta =\{\theta_0, \theta_{1}, \cdots, \theta_{L} \}$, whose element $\theta_{l}$ corresponds to $M_{l}$ for all $l \in \{0,\cdots,L \}$.
	The modulation mode table is updated to include the phase rotations, as shown in Table \ref{Table:ModePhaseExample}. 
	
	\begin{table}[t!]%
		\caption{Example of Modulation Mode Table for Phase-Rotated Modulation}
		\label{Table:ModePhaseExample}
		{\footnotesize
			\begin{center}
				\begin{tabular}{p{1.4cm}||c|c|p{0.3cm}|c}
					\toprule[1.2pt]
					Modulation Mode & $M_0$ & $M_{1}$ & $\cdots$ & $M_{L}$ \\
					\midrule
					Modulation orders & $(m_O,-)$ & $(m_{f,1},m_{n,1})$ & $\cdots$ & $(m_{f,L},m_{n,L})$\\
					\midrule
					Power coefficients & $(P,-)$ & $(P_{f,1},P_{n,1})$ & $\cdots$ & $(P_{f,L},P_{n,L})$\\
					\midrule
					Phase rotations & $\theta_0$ & $\theta_{1}$ & $\cdots$ & $\theta_{L}$\\
					\bottomrule[1.2pt]
				\end{tabular}
			\end{center}
		}
	\end{table}
	
	However, a larger $d_{\text{min}}(M_1e^{j\theta},M_2)$ does not guarantee better user capacity, because the SINR terms in (\ref{eq:capacity}) would be changed.
	We have mentioned the tradeoff between $p_{l \rightarrow m}$ and $\eta_{l \rightarrow m}$ in \eqref{eq:capacity} in Section \ref{sec:Capacity_analysis_for_blind_modulation_Detection}; although phase-rotated modulation can provide a lower classification error rate, the SINR degradation due to incorrect signal classification would be larger than that in legacy modulation.
	Therefore, we should carefully find the phase rotation values that maximize the user capacity by balancing the tradeoff between the classification error probability and the SINR degradation due to incorrect classification. 
	We can formulate the optimization problem to find the phase rotation list $\Theta$ as follows:
	
	\begin{equation}
	\Theta = \arg \max_{\Theta=\{\theta_0 , \cdots,\theta_{L}\}}{C}.
	\label{eq:optimization_Cmax}
	\end{equation}
	
	The above optimization problem is difficult to solve theoretically because the expectation in \eqref{eq:capacity} is taken over random channel realizations; we have obtained the optimal phase rotations numerically in this paper.
	When there exist many modulation modes, however, numerical determination of all the rotation values requires excessively massive computations.
	In this study, phase-rotated modulation was applied for only OMA/NOMA classification, i.e., $\theta_0 \neq 0$ and $\theta_{l}=0$ for all $l \in \{1,\cdots,L\}$, because incorrect OMA/NOMA classification results in zero throughput, as shown in Section \ref{sec:SINR_analysis}.
	However, incorrect determination of the far UT's modulation order or power allocation ratio is not as critical as OMA/NOMA misclassification.
	
	On the other hand, near/far UT classification is not affected by phase-rotated modulation.
	According to (\ref{ML_nearUT}) and (\ref{ML_farUT}), near/far UT classification depends on the constellation structures of $\chi_l$ and $\chi_l^f$.
	Phase-rotated modulation changes the constellations to $\chi_l e^{j\theta_{l}}$ and $\chi_l^f e^{j \theta_{l}}$, but the minimum distance between them is not changed.
	Hence, phase-rotated modulation cannot influence the performance of ML-based near/far UT classification.
	
	\begin{table}[t!]
		\caption{Characteristics of the proposed blind signal classification schemes}
		\label{table:BSC_comp}
		\begin{center}
			\begin{tabular}{|p{3cm}|c|c|}
				\hline
				{} & Phase-rotated & Pilot reuse-based \\
				\hline
				OMA/NOMA & \multirow{2}{*}{o} & \multirow{2}{*}{o} \\
				classification & & \\
				\hline
				Modulation & \multirow{2}{*}{o} & \multirow{2}{*}{o} \\
				classification & & \\
				\hline
				Near/Far UT & \multirow{2}{*}{x} & \multirow{2}{*}{o} \\
				classification & & \\
				\hline
				Additional overhead & x & x \\
				\hline
				Additional constraint & x & o \\
				\hline
				Complexity & high & low \\
				\hline
			\end{tabular}
		\end{center}
	\end{table}
	
	\section{Pilot Reuse-based Signal Classification}
	\label{sec:Pilot-Reuse_based_blind_OMA/NOMA_and_Modualtion_detection}
	
	In Section \ref{sec:phase_rotated_modulation}, the ML-based phase-rotated modulation scheme was proposed, which uses data symbols for blind signal classification, and we discussed the manner in which user capacity and classification accuracy change depending on the phase rotations. 
	However, because a tradeoff exists between the classification accuracy and SINR degradation in the case of a classification error, the ML-based phase-rotated modulation scheme is limited in terms of improving the blind classification performance.
	Therefore, a pilot-based scheme that does not affect the SINR is presented in this section; however, this scheme requires an additional constraint, as explained below.
	This signal classification algorithm based on pilot reuse requires no changes in the modulation scheme.
	In addition, phase-rotated modulation cannot improve near/far UT classification, which means that phase rotations of data symbols do not influence the near/far UT classification performance.
	Therefore, the phase rotations assigned to pilots are considered. 
	
\begin{figure} [t!]
		\centering
		\includegraphics[width=0.47\textwidth]{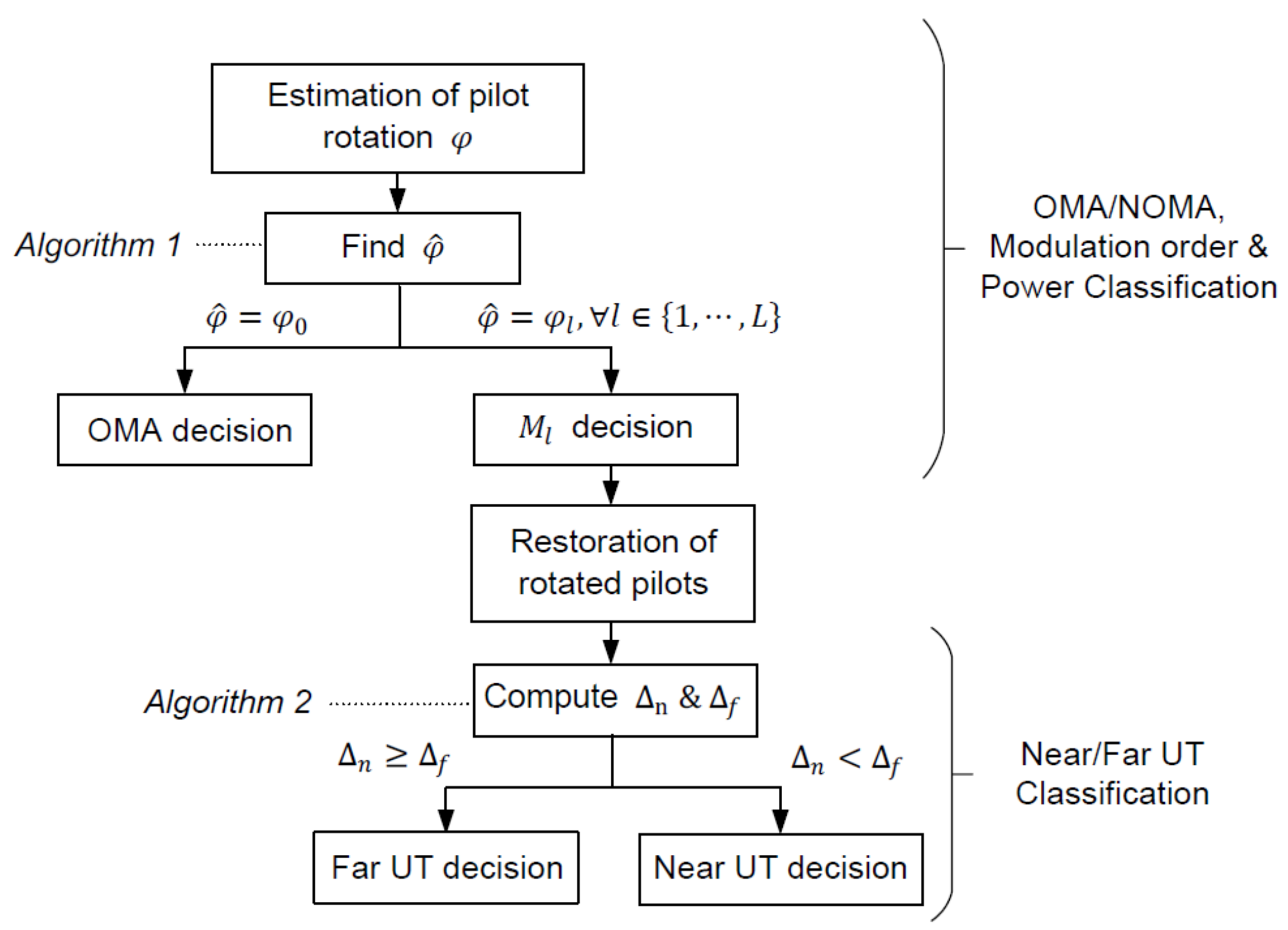}
		\caption{Processes of pilot reuse-based signal classification}
		\label{Fig:PRBD_scenario}
	\end{figure}

	The various processes of the pilot reuse-based signal classification algorithm are shown in Fig. \ref{Fig:PRBD_scenario}. 
	The OMA/NOMA classification and modulation order and power ratio selection are conducted simultaneously.
	Then, near/far UT classification is executed.
	A comparison of the ML-based phase rotated modulation and the pilot reuse-based signal classification scheme is briefly shown in Table \ref{table:BSC_comp}. 
	
	\subsection{Pilot Reuse-based OMA/NOMA and Modulation Classifications}
	
	In contrast to phase-rotated modulation, phase rotations for existing pilots are introduced in this subsection.
	The data symbols are conventionally modulated; therefore, the SINR terms in (\ref{eq:capacity}) do not change.
	This algorithm does not require additional pilots.
	It just rotates the existing pilot symbols already used for other purposes, e.g., carrier frequency offset estimation.
	
	The proposed pilot reuse-based scheme requires some assumptions.
	First, the channel should be static for at least two pilot symbol durations.
	Second, an identical value should be used for two consecutive pilots that experience the static channel gain.
	In the proposed method, the BS transmits the legacy value during the first pilot symbol duration and rotates the identical pilot during the second pilot symbol duration. 
	Different phase rotations are assigned to each modulation mode, as in phase-rotated modulation. 
	The symbol $\phi$ is used to denote the pilot rotation in order to avoid confusion with $\theta$ in phase-rotated modulation.
	
	Assuming $M_{l_0}$ is used to modulate the signal, the two received consecutive pilot symbols are given by
	\begin{align}
	r_u &= hp_u + w_u
	\label{eq:legacyPilot} \\
	r_r &= hp_r+n_r = hp_u e^{j\phi_{l_0}} + w_r,
	\label{eq:rotatedPilot}
	\end{align}
	where $p$ and $r$ are the transmitted and received pilots, respectively.
	The subscripts $\mathit{u}$ and $\mathit{r}$ denote unrotated and rotated symbols, respectively.
	In addition, $w_u, w_r \sim \mathcal{CN}(0,\sigma_n^2)$.
	The receiver can estimate the phase rotation of the pilot in the second symbol duration as
	\begin{equation}
	\varphi = \measuredangle\{(r_{u})^* r_{r}\} \approx \measuredangle \{|r_{u}|^2 e^{j\phi_{l_0}}\}.
	\label{eq:thetaEstimation1}
	\end{equation}
	By comparing the estimated $\varphi$ with the exact rotations, the modulation mode, $M_{\hat{l}}$, can be easily classified as
	\begin{equation}
	\hat{l} = \underset{l\in \{0,\cdots,L,\}}{\arg\min}~|\varphi-\phi_{l}|.
	\label{eq:PRC_decision}
	\end{equation}
	
	Since different phase rotation values are assigned to each modulation mode, the determination of the transmitted modulation mode in \eqref{eq:PRC_decision} can be interpreted as finding the decision region of a particular modulation mode in which the estimated phase $\varphi$ is placed.
	In other words, we can make $L+1$ exclusive regions in $[0,2\pi)$ and assign separate regions to all modulation modes as their decision regions.
	First, let $\boldsymbol{\Phi}_O$ and $\boldsymbol{\Phi}_N$ be the phase ranges for OMA and NOMA with different intervals in $[0,2\pi)$, respectively. 
	Then, obviously, $\phi_0 \in \boldsymbol{\Phi}_O$ and $\phi_{1},\cdots,\phi_{L} \in \boldsymbol{\Phi}_N$. 
	It is also assumed that $\boldsymbol{\Phi}_O \cup \boldsymbol{\Phi}_N = [0,2\pi)$ and $\boldsymbol{\Phi}_O \cap \boldsymbol{\Phi}_N = \{\phi \}$, where $\{\phi\}$ is the empty set.
	The estimated $\varphi$ must be included in either $\boldsymbol{\Phi}_O$ or $\boldsymbol{\Phi}_N$, but not in both.
	If $\varphi \in \boldsymbol{\Phi}_O$, the signal is classified as OMA; otherwise, as NOMA.
	Accordingly, $\boldsymbol{\Phi}_O$ and $\boldsymbol{\Phi}_N$ are the decision regions of OMA and NOMA, respectively.
	
	Similarly, the far UT's modulation order can be classified by dividing $\boldsymbol{\Phi}_N$ into several nonoverlapping phase ranges corresponding to different far UT's modulation order candidates.
	For example, let QAM, 16-QAM, and 64-QAM be the candidates of a far UT's modulation.
	Then, we can generate $\boldsymbol{\Phi}_{N}^{QAM}$, $\boldsymbol{\Phi}_N^{16QAM}$, and $\boldsymbol{\Phi}_N^{64QAM}$, corresponding to the NOMA modes using QAM, 16-QAM, and 64-QAM as the far UT's modulation, respectively.
	These ranges satisfy $\boldsymbol{\Phi}_{N}^{QAM} \cup \boldsymbol{\Phi}_N^{16QAM} \cup \boldsymbol{\Phi}_N^{64QAM} = \boldsymbol{\Phi}_N$.
	Therefore, if $\varphi \in \boldsymbol{\Phi}_N$, then $\varphi$ must be included in only one range among $\boldsymbol{\Phi}_{N}^{QAM}$, $\boldsymbol{\Phi}_N^{16QAM}$, and $\boldsymbol{\Phi}_N^{64QAM}$, and the far UT's modulation can be determined.
	
	Next, suppose that QAM is classified as the far UT's modulation and there exist $L_1$ modulation modes with different power weightings. 
	Again, we can generate decision regions for the classification of the power ratio by separating $\boldsymbol{\Phi}_N^{QAM}$ into $L_1$ nonoverlapping ranges denoted by $\boldsymbol{\Phi}_{N,l}^{QAM}$ for all $l\in \{1,\cdots,L_1\}$, satisfying $\bigcup_{l=1}^{L_1} \Phi_{N,l}^{QAM} = \Phi_N^{QAM}$, similar to the aforementioned classification steps. 
	A series of the pilot reuse-based classification processes of OMA/NOMA and modulation are expressed in Algorithm \ref{alg:PRBD}.
	In Algorithm \ref{alg:PRBD}, we assume that the far UT's modulation of the first $L_1$ modes, $M_{1},...M_{L_1}$, is QAM. 
	In addition, the next $L_2$ modes, $M_{L_1+1},...,M_{L_1+L_2}$, and the remaining $L_3$ modes, $M_{L_1+L_2+1},...M_{L}$, use 16-QAM and 64-QAM for the far UT's modulation, respectively.
	
	\begin{algorithm}[t!]
		\caption{Pilot Reuse-based OMA/NOMA and Modulation Classifications
			\label{alg:PRBD}}
		\begin{algorithmic}[1]
			\Require{Phase ranges: \newline$\boldsymbol{\Phi}_O , \boldsymbol{\Phi}_N , \boldsymbol{\Phi}_{N}^{QAM}, \boldsymbol{\Phi}_N^{16QAM}, \boldsymbol{\Phi}_N^{64QAM}$}
			\State {Compute $\varphi$ by (\ref{eq:thetaEstimation1})}
			\If {$\varphi \in \boldsymbol{\Phi}_0$}
			\State {Decide $M_0$}
			\Else
			\If {$\varphi \in \boldsymbol{\Phi}_N^{QAM}$}
			\Let {$\hat{l}$} {$\underset{l=1,...,L_1}{\arg\min} |\phi_{l} - \varphi|$}
			\ElsIf {$\varphi \in \boldsymbol{\Phi}_N^{16QAM}$}
			\Let {$\hat{l}$} {$\underset{l=L_1+1,...,L_1+L_2}{\arg\min} |\phi_{l} - \varphi|$}
			\Else
			\Let {$\hat{l}$} {$\underset{l=L_1+L_2+1,...,L}{\arg\min} |\phi_{l} - \varphi|$}
			\EndIf
			\State {Decide $M_{\hat{l}}$}
			\EndIf
		\end{algorithmic}
	\end{algorithm}
	
	Algorithm \ref{alg:PRBD} performs OMA/NOMA and modulation classifications simultaneously; therefore, blind signal classification becomes simpler.
	In addition, this scheme does not require the rotation value to be accurately estimated. 
	It is sufficient that the roughly estimated $\varphi$ is in the decision region of the transmitted modulation mode for correct signal classification. 
	After MC, the pilot symbols should serve their original purposes; therefore, the rotated symbols must be de-rotated, $r_r e^{-j\phi_{\hat{l}}}$.
	In this case, even though the estimation of $\varphi$ is not accurate, if MC is correct, then we can find the accurate phase rotation value $\phi_{\hat{l}}$ according to the modulation mode table, which is shown in Table \ref{Table:ModePhaseExample}.
	Thus, the proposed classification algorithm does not need to know the pilot value and the channel gain. 
	
	The performance of the proposed algorithm depends on the method of determining the phase rotation values and decision regions for each modulation mode. 
	From a broad perspective, there are two phase assignment rules as follows.
	
	\subsubsection{Uniform Assignment}
	The simplest rule is the uniform one, $\phi_l = \frac{2\pi \cdot l}{L+1}$ for $M_{l}$ and $\phi_O = 0$ for $M_O$.
	In this case, $\boldsymbol{\Phi}_N = [\frac{2\pi}{L+1},2\pi-\frac{2\pi}{L+1})$ and $\boldsymbol{\Phi}_O = [0,\frac{2\pi}{L+1}) \cup [2\pi-\frac{2\pi}{L+1},2\pi)$.
	The uniform assignment rule seems reasonable; however, $\boldsymbol{\Phi}_O$ becomes smaller as $L$ increases.
	Therefore, it is unfair when an OMA signal is transmitted.
	
	\subsubsection{Non-uniform Assignment}
	We can give greater importance to OMA/NOMA classification than to the classification of modulation order or power ratio by using the non-uniform assignments of phase rotations. 
	Because OMA/NOMA classification is more important than MC, the generation of $\boldsymbol{\Phi}_O$ and $\boldsymbol{\Phi}_N$ have the first priority, followed by $\boldsymbol{\Phi}_N^{QAM}$, $\boldsymbol{\Phi}_N^{16QAM}$, and $\boldsymbol{\Phi}_N^{64QAM}$.
	Finally, the phase rotations of the NOMA modes having the same far UT's modulation order but different power ratios are determined.
	
	\begin{figure} [t!]
		\centering
		\includegraphics[width=0.35\textwidth]{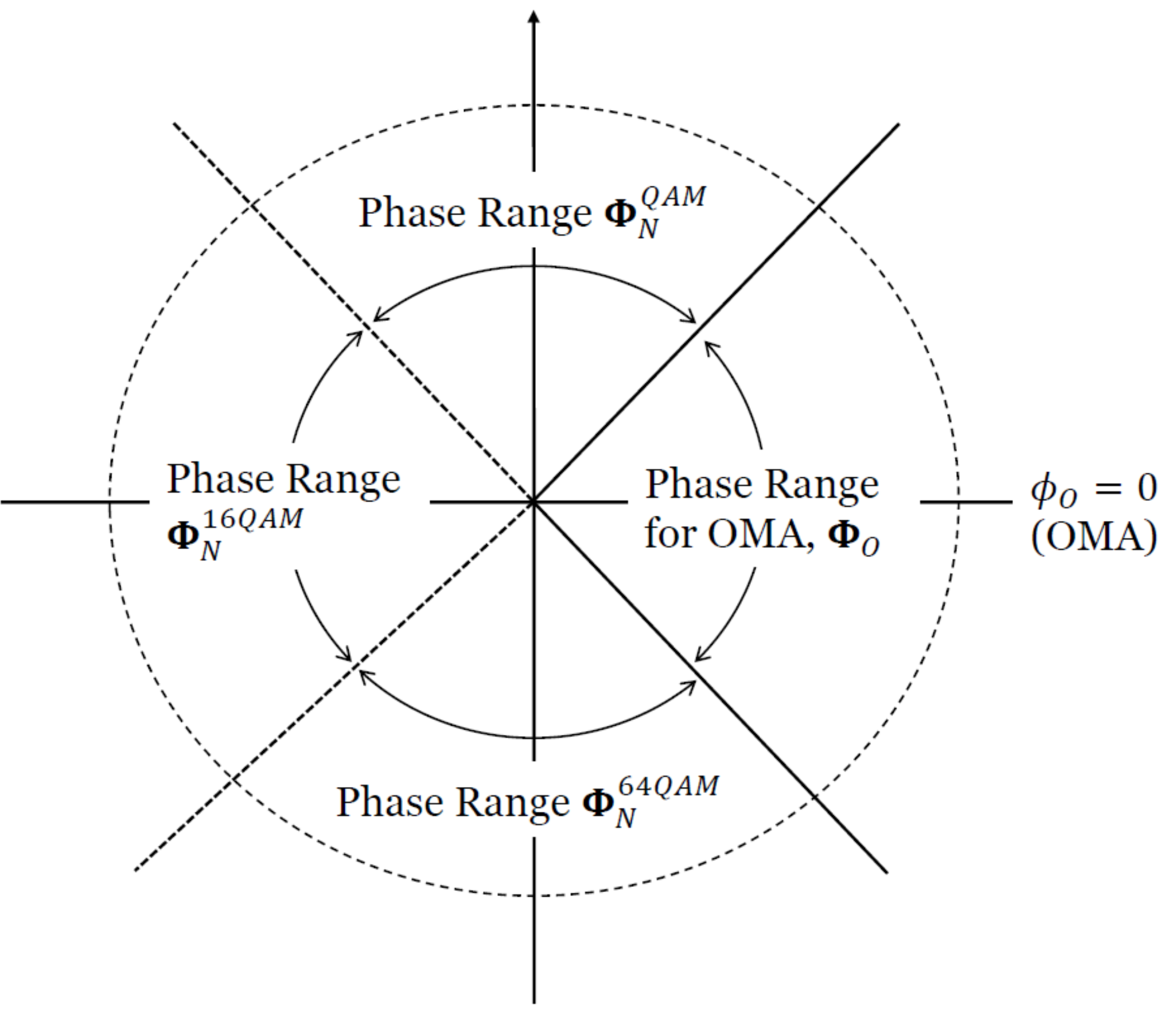}
		\caption{Phase ranges of non-uniform phase rotation rule}
		\label{Fig:PhaseRule}
	\end{figure}
	
	The phase ranges of the non-uniform assignments are shown in Fig. \ref{Fig:PhaseRule}.
	Although $\mathbf{\Phi}_N^{QAM}$, $\mathbf{\Phi}_N^{16QAM}$, and $\mathbf{\Phi}_N^{64QAM}$ consist of $L_1$, $L_2$, and $L_3$ modes, these ranges occupy the same amount of interval as that of $\mathbf{\Phi}_O$.
	After $\boldsymbol{\Phi}_O$ and $\boldsymbol{\Phi}_N$ are generated, $\boldsymbol{\Phi}_N$ is divided into $\boldsymbol{\Phi}_N^{QAM}$, $\boldsymbol{\Phi}_N^{16QAM}$, and $\boldsymbol{\Phi}_N^{64QAM}$, as shown in Fig. \ref{Fig:PhaseRule}.
	The interval sizes of the phase ranges can be arbitrarily chosen depending on the parameters of the system environment, such as the number of modulation modes.
	The exact phase rotation values for the NOMA modes with different power weightings, i.e., $\phi_{1},\cdots,\phi_{L}$, can be uniformly chosen in the range in which each mode is included, because power ratio classification is less important than the other classification steps.
	For example, if $\boldsymbol{\Phi}_N^{QAM}=[\frac{\pi}{4}, \frac{3\pi}{4})$, then $\phi_{l} = \frac{\pi}{4} + \frac{\pi}{2L_1} \cdot (l-1)$, for all $l \in \{1,...,L_1\}$.
	
	\subsection{Pilot Reuse-based Near/Far UT Classification}
	\label{sec:Proposed_Blind_Near/Far_UT_Detection}
	
	Since the two signals with different power weightings are transmitted simultaneously, the pilot symbols targeting the two UTs are also superposed and transmitted. 
	When the transmitted signal is modulated by $M_l$, the power-multiplexed legacy pilot becomes
	\begin{equation}
	p_l = \sqrt{P_{f}}p_l^f + \sqrt{P_{n}}p_l^n,
	\label{eq:powermultiplexed_pilot}
	\end{equation}
	where $p_l^f$ and $p_l^n$ are legacy pilots for the far and near UTs, respectively.
	The rotated pilot symbols used for OMA/NOMA and modulation classifications can also be utilized for near/far UT classification after they have been de-rotated.
	Even though each UT knows only its own pilot values, the receiver cannot recognize which power ratio is weighted to its pilot symbol because it does not perform near/far UT classification yet.
	Let $p_l^0$ be a known pilot value. 
	Then, if the receiver is far UT, $p_l^0 = p_l^f$, otherwise, $p_l^0 = p_l^n$. 
	
	The proposed near/far UT classification algorithm in a two-user NOMA system is summarized in Algorithm \ref{alg:Far/NearUT}.
	This algorithm requires that channel estimation and MC should have been completed in advance.
	When the MC correctly determines $M_{l}$, the UT computes two hypotheses, $\Delta_l^f$ and $\Delta_l^n$, each of which is true when the receiver is the far or near UT, respectively. 
	
	\begin{algorithm}[t!]
		\caption{Pilot Reuse-based Near/Far UT Classification
			\label{alg:Far/NearUT}}
		\begin{algorithmic}[1]
			\Require{$p_{l}^0$, $\chi_l$, $\chi_l^f$}
			\State {Compute $a_f$ and $a_n$}
			\State \indent {$a_f = y - \hat{h} \sqrt{P_f} p_{l}^0$}
			\State \indent {$a_n = y - \hat{h} \sqrt{P_n} p_{l}^0$}
			\State {Compute $\Delta_l^f$ and $\Delta_l^n$}
			\State \indent {$\Delta_l^f = \min_{q \in \chi_l}|a_f - \hat{h}\sqrt{P_n}q|$}
			\State \indent {$\Delta_l^n = \min_{q \in \chi_l^f}|a_n - \hat{h}\sqrt{P_f}q|$}
			\If {$\Delta_l^f \geq \Delta_l^n$}
			\State {Near UT Decision}
			\Else
			\State {Far UT Decision}
			\EndIf
		\end{algorithmic}
	\end{algorithm}
	
	To explain the algorithm clearly, an example is presented. 
	From (\ref{eq:legacyPilot}) and (\ref{eq:powermultiplexed_pilot}), the received pilot is given by
	\begin{equation}
	y_{l} = h(\sqrt{P_f}p_l^f + \sqrt{P_n}p_l^n) + n_l.
	\label{eq:ex:received}
	\end{equation}
	Suppose that the channel estimation is perfect and the receiver is the near UT, i.e., $p_l^0 = p_l^n$; then,
	\begin{align}
	a_f &= y_{l} - h\sqrt{P_f}p_{l}^0 = h(\sqrt{P_f}(p_{l}^f-p_l^n) + \sqrt{P_n}p_l^n) + n_{l}
	\label{eq:ex:a_1^f}\\
	a_n &= y_{l} - h\sqrt{P_n}p_{l}^0 = h\sqrt{P_f}p_{l}^f + n_l.
	\label{eq:ex:a_1^n}
	\end{align}
	Here, $a_f$ and $a_n$ are obtained under the assumption that the receiver would be a far and near UT, respectively.
	The next step is to compute $\Delta_l^f$ and $\Delta_l^n$ as follows:
	\begin{align}
	\Delta_l^f &= \min_{q \in \chi_l}|a_f - h\sqrt{P_n}q|
	\label{eq:ex:Delta_1^f} \\
	\Delta_l^n &= \min_{q \in \chi_l^f}|a_n - h\sqrt{P_f}q|.
	\label{eq:ex:Delta_1^n}
	\end{align}
	
	Because MC has been completed, $P_f$, $P_n$, $\chi_l^f$, and $\chi_l$ are known.
	Note that $\Delta_l^n$ remains the only noise component when $q=p_{l}^f$.
	However, $a_f$ includes the non-zero term in (\ref{eq:ex:a_1^f}), i.e., $h\sqrt{P_f}(p_{l}^f-p_l^n)$, as well as the noise component. 
	Thus, mostly $\Delta_l^f \geq \Delta_l^n$ and the near UT is determined.
	Otherwise, the receiver does not require SIC, i.e., it is determined as the far UT. 
	
	\section{Performance Evaluation}
	\label{sec:SimulationResults}
	
	\subsection{Simulation Environments}
	
	This section describes a variety of performance comparisons of conventional ML signal classification with the proposed schemes. Acronyms are used for the signal classification methods in the related figures: ``MLC" for ML classifier, ``MLC-PRM" for phase-rotated modulation based on ML classification, and ``PRC" for the pilot reuse-based classifier.
	For MLC and MLC-PRM, 10 data symbols were used to classify the received signals.
	PRC utilized only one pair of pilots, because the number of pilots is usually smaller than that of data symbols.
	Three example cases of the modulation modes are presented.
	
	\begin{table}[t!]%
		\caption{Case 1: Modulation Mode Table}
		\label{Table:ModeExample1}
		{\footnotesize
			\begin{center}
				\begin{tabular}{c|c|c|c}
					\toprule[1.2pt]
					Modulation & Modulation & Modulation & Power ratio \\
					mode & (far UT) & (near UT) & (far UT) \\
					\midrule
					$M_0$ & QPSK & - & 1.0 \\
					\midrule
					$M_{1}$ & QPSK & QPSK & 0.8 \\
					\midrule
					$M_{2}$ & QPSK & QPSK & 0.8621 \\
					\midrule
					$M_{3}$ & QPSK & QPSK & 0.9163 \\
					\bottomrule[1.2pt]
				\end{tabular}
			\end{center}
		}
	\end{table}
	
	\begin{table}[t!]%
		\caption{Case 2: Modulation Mode Table}
		\label{Table:ModeExample2}
		{\footnotesize
			\begin{center}
				\begin{tabular}{c|c|c|c}
					\toprule[1.2pt]
					Modulation & Modulation & Modulation & Power ratio \\
					mode & (far UT) & (near UT) & (far UT) \\
					\midrule
					$M_0$ & 16-QAM & - & 1.0 \\
					\midrule
					$M_{1}$ & QPSK & 16-QAM & 0.8653 \\
					\midrule
					$M_{2}$ & 16-QAM & 16-QAM & 0.95 \\
					\bottomrule[1.2pt]
				\end{tabular}
			\end{center}
		}
	\end{table}
	
	\begin{table}[t!]%
		\caption{Case 3: Modulation Mode Table}
		\label{Table:ModeExample3}
		{\footnotesize
			\begin{center}
				\begin{tabular}{c|c|c|c}
					\toprule[1.2pt]
					Modulation & Modulation & Modulation & Power ratio \\
					mode & (far UT) & (near UT) & (far UT) \\
					\midrule
					$M_0$ & 16-QAM & - & 1.0 \\
					\midrule
					$M_{1}$ & QPSK & 16-QAM & 0.7619 \\
					\midrule
					$M_{2}$ & QPSK & 16-QAM & 0.8653 \\
					\midrule
					$M_{3}$ & QPSK & 16-QAM & 0.9275 \\
					\midrule
					$M_{4}$ & 16-QAM & 16-QAM & 0.95 \\
					\midrule
					$M_{5}$ & 16-QAM & 16-QAM & 0.97 \\
					\bottomrule[1.2pt]
				\end{tabular}
			\end{center}
		}
	\end{table}
	
	\subsubsection{Case 1}
	Case 1 is based on Table \ref{Table:ModeExample1}. 
	The far UT's modulation is fixed; therefore, OMA/NOMA and power ratio classifications are considered.
	It is supposed that $M_{2}$ is transmitted.
	
	\subsubsection{Case 2}
	Case 2 is based on Table \ref{Table:ModeExample2}. 
	A single power ratio is assigned to each mode; therefore, OMA/NOMA and the far UT's modulation order classifications are considered.
	It is supposed that $M_{1}$ is transmitted.
	
	\subsubsection{Case 3}
	Case 3 is based on Table \ref{Table:ModeExample3}. 
	It considers OMA/NOMA, the power ratio, and the far UT's modulation order classifications.
	It is supposed that $M_{2}$ is transmitted.
	
	The power ratios of the modulation modes, in which the far UT's signal is modulated by QPSK in Tables \ref{Table:ModeExample1}-\ref{Table:ModeExample3}, follow the MuST parameters of 3GPP \cite{MuST:3GPP}.
	Since MuST considers QPSK only for the far UT from this point onward, the power ratios of the far UT that uses 16-QAM were arbitrarily chosen.
	
	We considered a two-user cellular NOMA system assuming a Rayleigh fading channel, $h \sim CN(0,1)$. 
	An ML equalizer and a low-density parity check (LDPC) 11ad decoder \cite{ldpc_11ad} were used for word error rate (WER) simulations.
	CWIC was basically used for SIC, but the system occasionally chose SLIC when the decoder-feedback could not be obtained, i.e., the classification of the far UT's modulation order was incorrect.
	The phase rotations of MLC-PRM were $\theta_0=0.6, 0.51$, and $0.69$ radians optimized at 13 dB, 20 dB, and 20 dB of SNR in Cases 1, 2, and 3, respectively.
	As explained previously, these phase rotations are applied for only OMA/NOMA classification.
	Additionally, the uniform phase assignment rule was used for PRC.
	
	\begin{figure} [t!]
		\centering
		\includegraphics[width=0.3\textwidth]{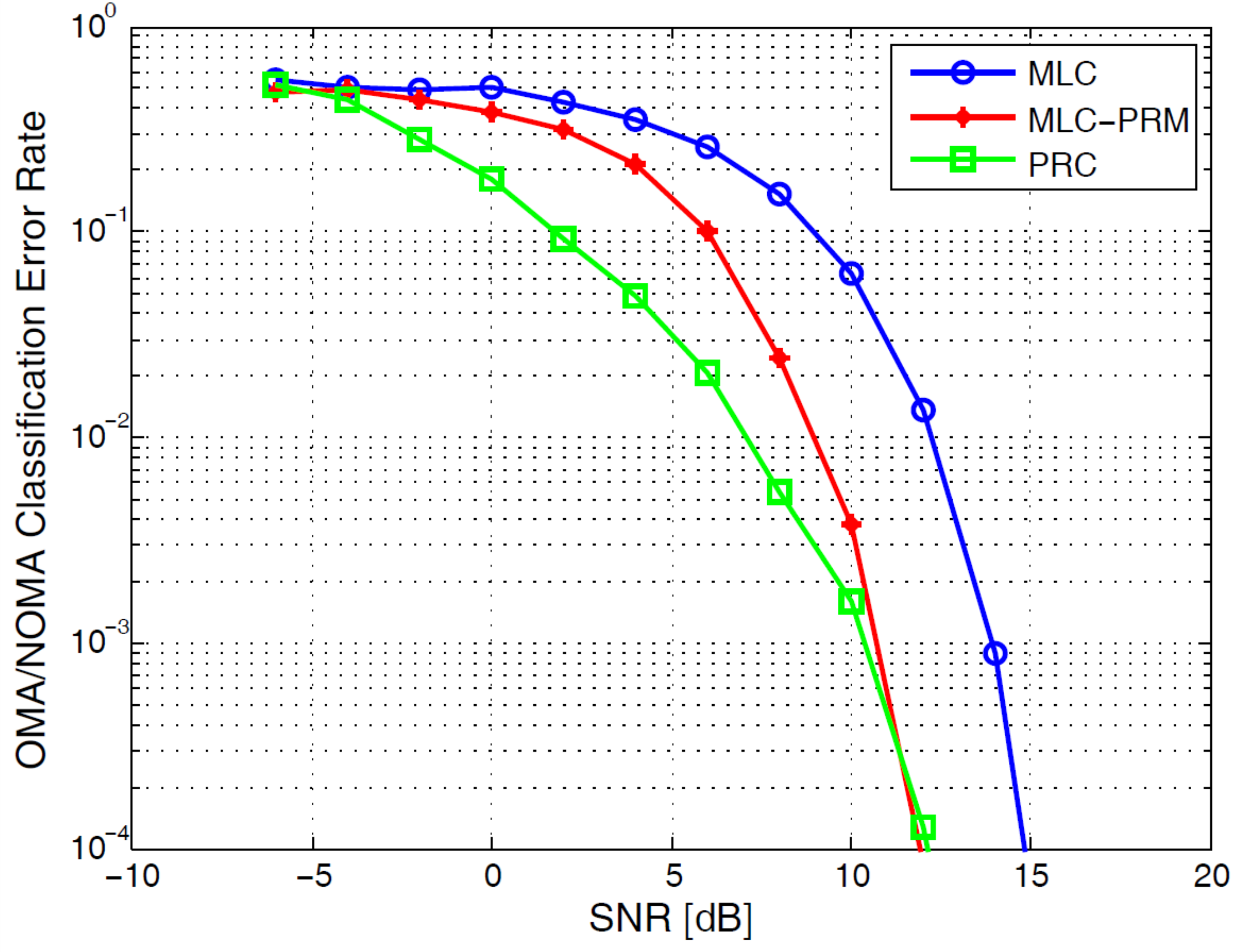}
		\caption{OMA/NOMA classification error rates in Case 1}
		\label{Fig:DetOMANOMAErr_Ex1}
	\end{figure}
	\begin{figure} [t!]
		\centering
		\includegraphics[width=0.3\textwidth]{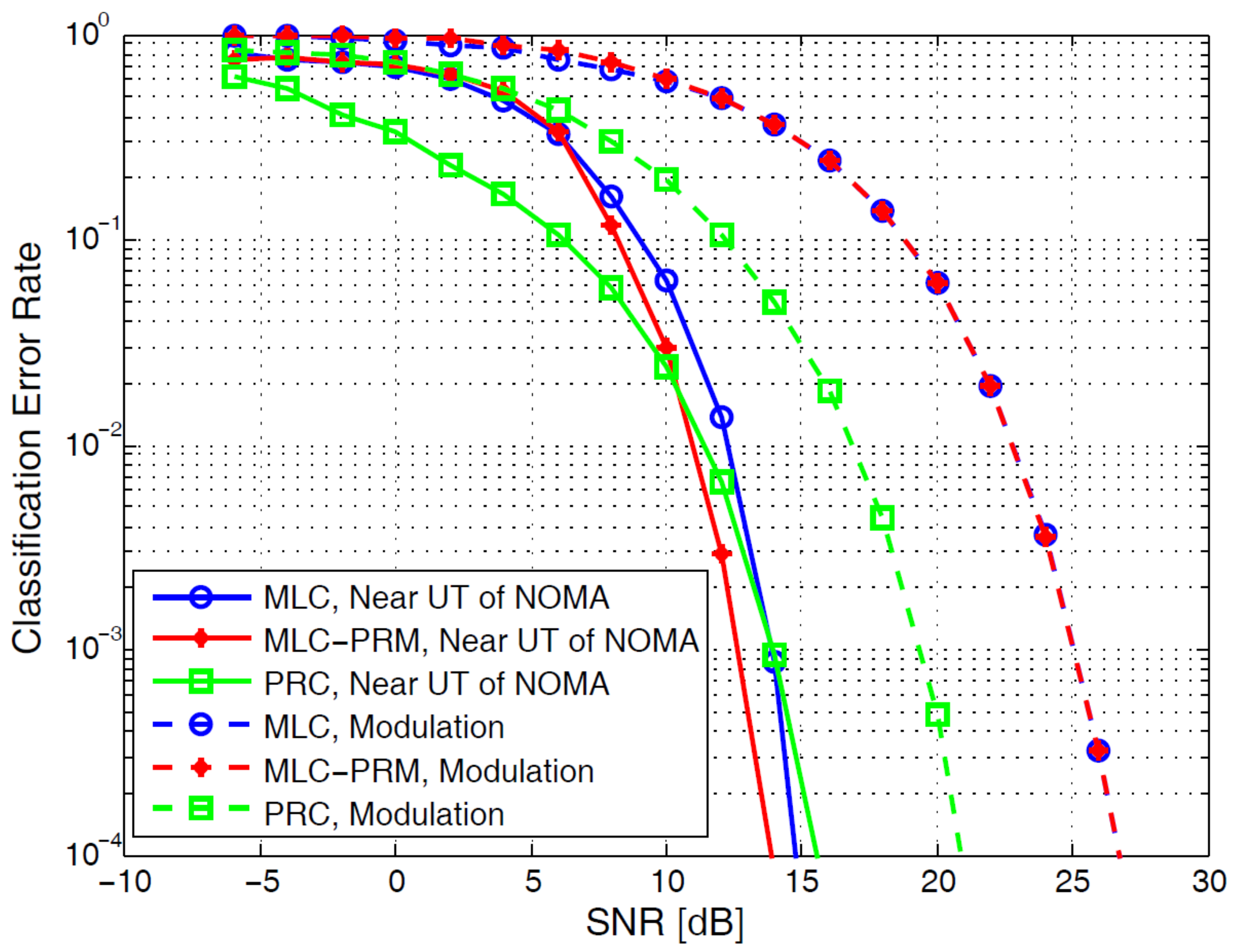}
		\caption{Near/far UT and modulation classification error rates in Case 1}
		\label{Fig:DetNOMANearErr_Ex1}
	\end{figure}
	\begin{figure} [t!]
		\centering
		\includegraphics[width=0.3\textwidth]{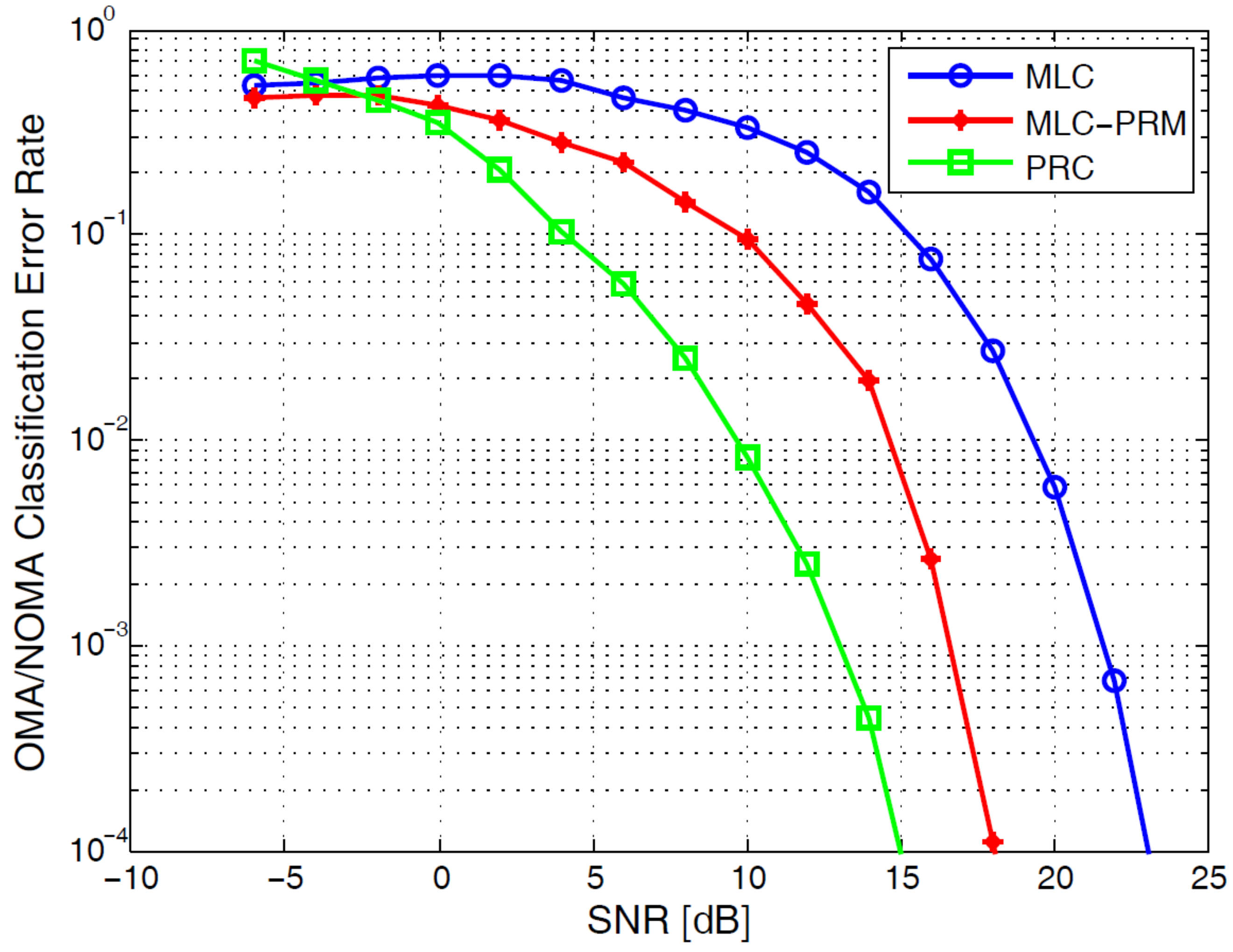}
		\caption{OMA/NOMA classification error rates in Case 2}
		\label{Fig:DetOMANOMAErr_Ex2}
	\end{figure}
	
	\begin{figure} [t!]
		\centering
		\includegraphics[width=0.3\textwidth]{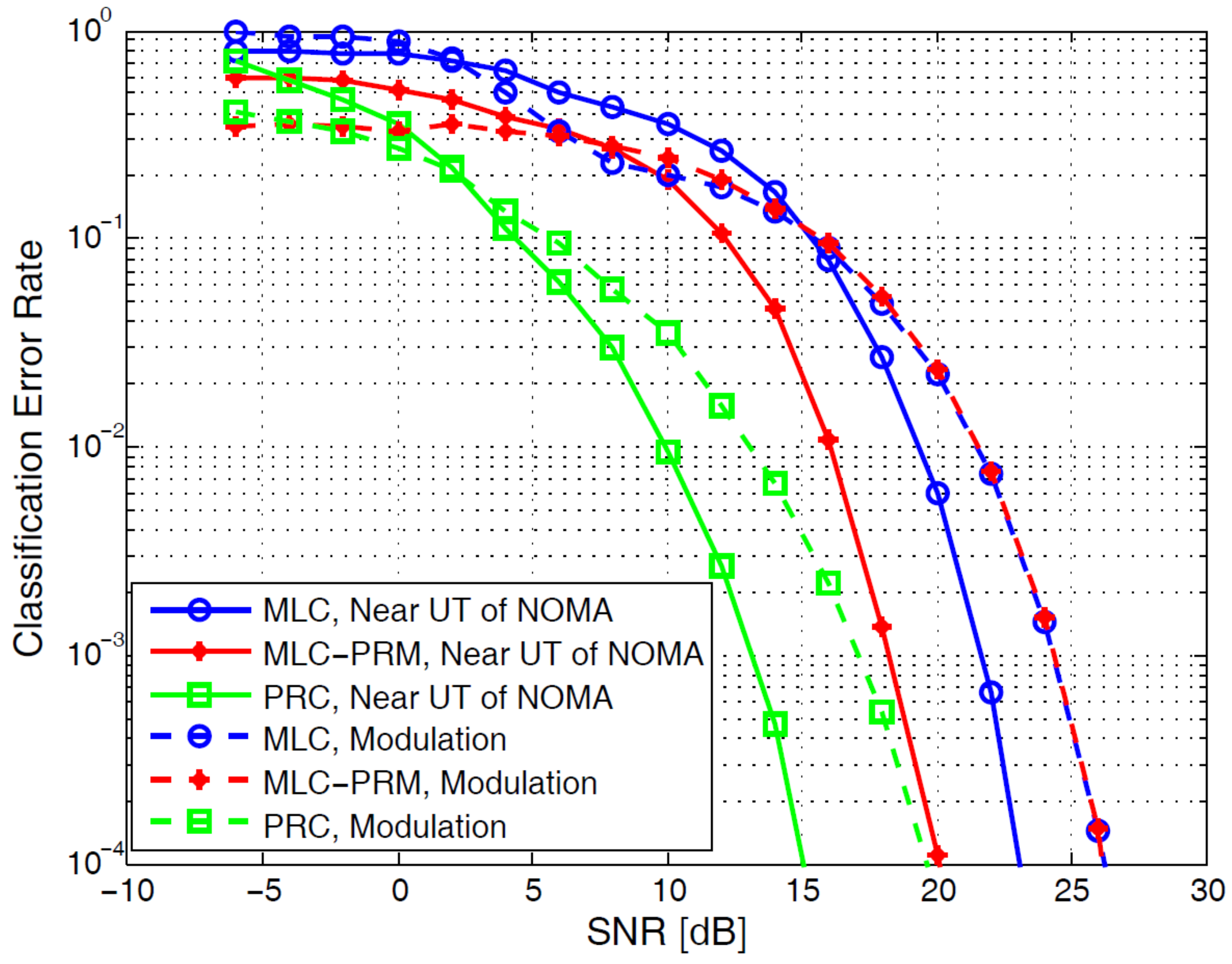}
		\caption{Near/far UT and modulation classification error rates in Case 2}
		\label{Fig:DetNOMANearErr_Ex2}
	\end{figure}
	
	\begin{figure} [t!]
		\centering
		\includegraphics[width=0.3\textwidth]{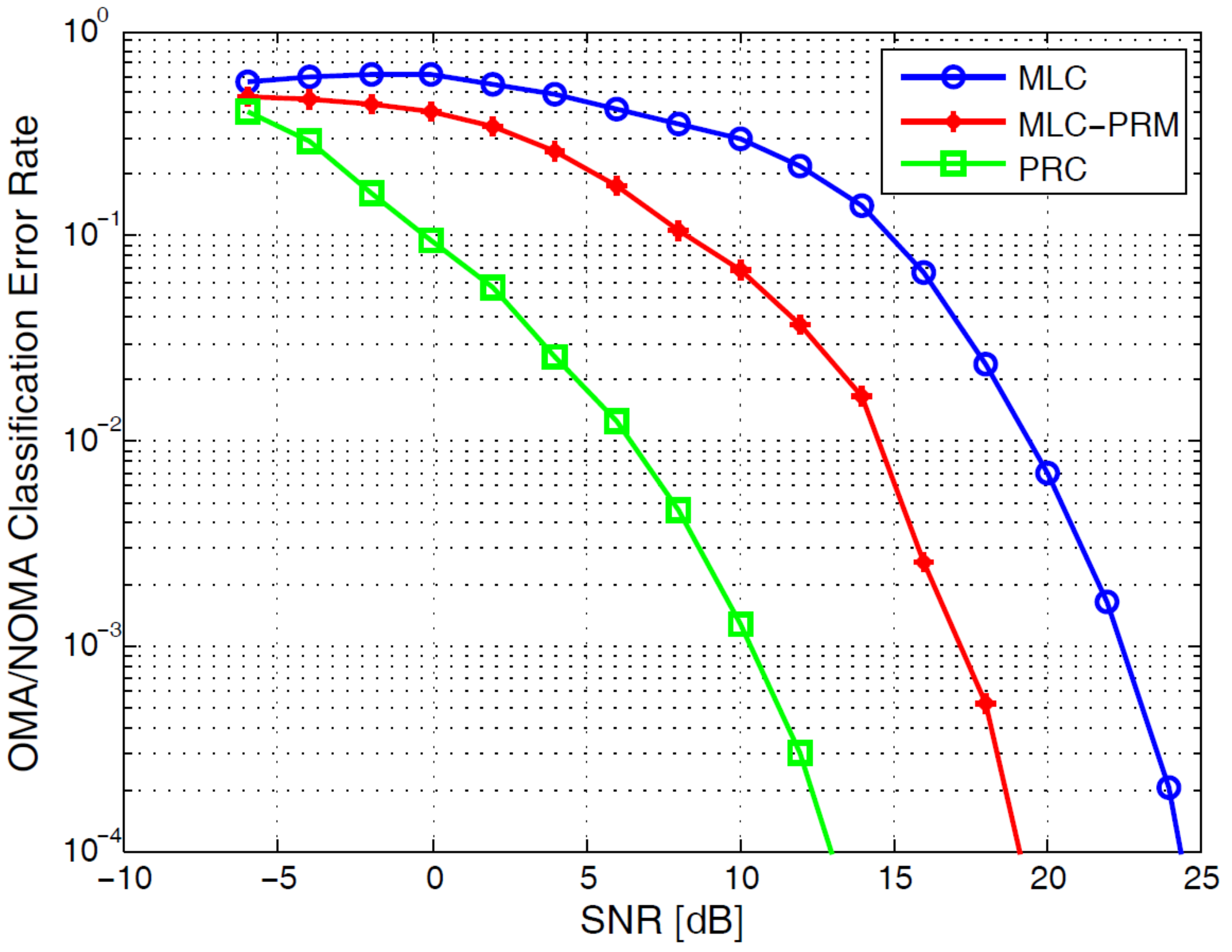}
		\caption{OMA/NOMA classification error rates in Case 3}
		\label{Fig:DetOMANOMAErr_Ex3}
	\end{figure}
	\begin{figure} [t!]
		\centering
		\includegraphics[width=0.3\textwidth]{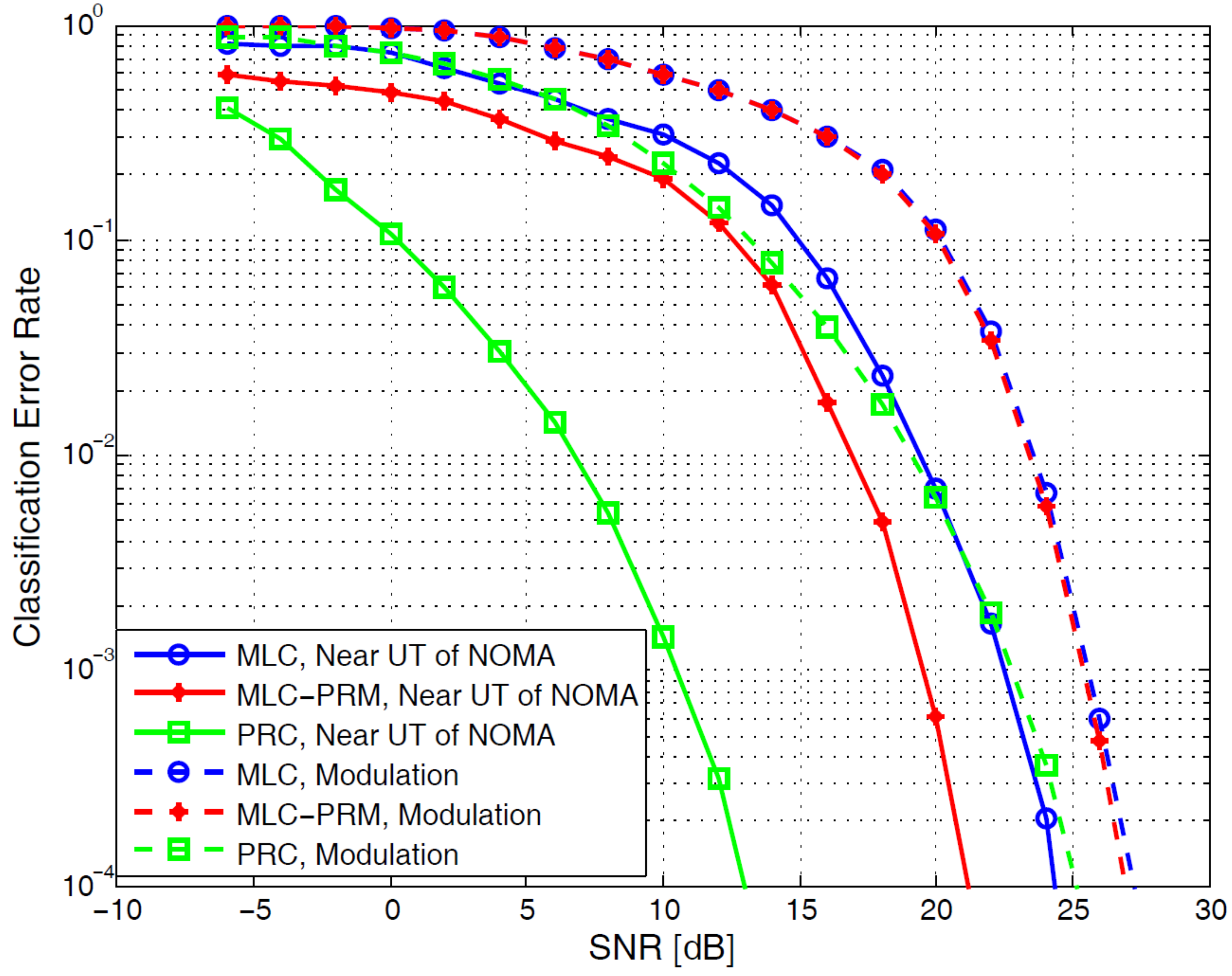}
		\caption{Near/far UT and modulation classification error rates in Case 3}
		\label{Fig:DetNOMANearErr_Ex3}
	\end{figure}
	
	\subsection{Signal Classification Error Rates and User Capacity}
	
	Figs. \ref{Fig:DetOMANOMAErr_Ex1}, \ref{Fig:DetOMANOMAErr_Ex2}, and \ref{Fig:DetOMANOMAErr_Ex3} show the OMA classification error rates in Cases 1, 2, and 3, respectively.
	We can easily see that PRC yields considerably better performances than MLC and MLC-PRM in each case. 
	MLC-PRM is also clearly better than MLC, but not as good as PRC.
	Since an OMA/NOMA classification error hampers correct data restoration, the proposed MLC-PRM and PRC are expected to be favorable for data detection.
	
	The near/far UT classification error rates in Cases 1, 2, and 3 are represented by the solid curves in Figs. \ref{Fig:DetNOMANearErr_Ex1}, \ref{Fig:DetNOMANearErr_Ex2}, and
	\ref{Fig:DetNOMANearErr_Ex3}, respectively.
	Because the signals determined as OMA do not require near/far UT classification, these error rate curves include incorrect determinations of OMA, as well as of far UT of NOMA.
	As compared to OMA/NOMA classification, the performance improvements of MLC-PRM over MLC in near/far UT classification are reduced.
	It means that even though MLC-PRM correctly classifies the signals that are significantly contaminated by noise or channel fading as NOMA, they fail to be determined as signals for the near UT in the next step because MLC-PRM does not improve near/far UT classification, as explained in Section \ref{sec:phase_rotated_modulation}. 
	In MLC, these signals have been already classified as OMA in the first step (i.e., OMA/NOMA classification); therefore, they do not exacerbate the near/far UT classification error rate as much as MLC-PRM.
	A similar phenomenon can be seen in the PRC graphs, but the decrease in performance gain of near/far UT classification is not significant, because PRC is designed to improve both OMA/NOMA and near/far UT classification steps.
	
	A comparison of the figures of the simulation cases reveals that there is not much difference in the near UT classification rates of the three methods shown in Fig. \ref{Fig:DetNOMANearErr_Ex1} as compared to those shown in Figs. \ref{Fig:DetNOMANearErr_Ex2} and \ref{Fig:DetNOMANearErr_Ex3}.
	In particular, the performances of MLC and MLC-PRM are similar to that of PRC in the high SNR region of Case 1. 
	This is because Case 2 and Case 3 are less sensitive to near/far UT classification than Case 1.
	In Case 1, every NOMA mode has the same modulation order for both UTs in Case 1; therefore, even if the signal is classified as NOMA, near/far UT classification should always be performed for correct data detection.
	However, in Cases 2 and 3, there are some situations where near/far UT classification is not required; in other words, the modulation orders for near and far UTs are different for some modes. 
	Note that the receiver already knows its own modulation order.
	
	The dashed curves in Figs. \ref{Fig:DetNOMANearErr_Ex1}, \ref{Fig:DetNOMANearErr_Ex2}, and
	\ref{Fig:DetNOMANearErr_Ex3} represent the MC error rates among the signals correctly classified as near UT of NOMA.
	Incorrect OMA/NOMA or near/far UT classification almost always results in a packet error; therefore, only the MC error rates of the signals determined to be near UT of NOMA are meaningful.
	The MC error rates look better than the near/far UT classification error rates in the low SNR region, because most of the signals severely contaminated by noise or channel fading are already incorrectly classified as OMA or the far UT of NOMA; therefore, these signals do not influence the MC error rates.
	However, since the modulation classification step compares much more hypotheses than OMA/NOMA and near/far UT classification, the MC error rates become worse than the near/far UT classification error rates as SNR grows. 
	In the case of MLC-PRM, a data symbol rotation is applied for only OMA/NOMA classification, and therefore, the MC error rates of MLC and MLC-PRM are almost the same in each case.
	However, PRC shows considerably better MC performances than both MLC and MLC-PRM.
	
	\begin{figure} [t!]
		\centering
		\includegraphics[width=0.3\textwidth]{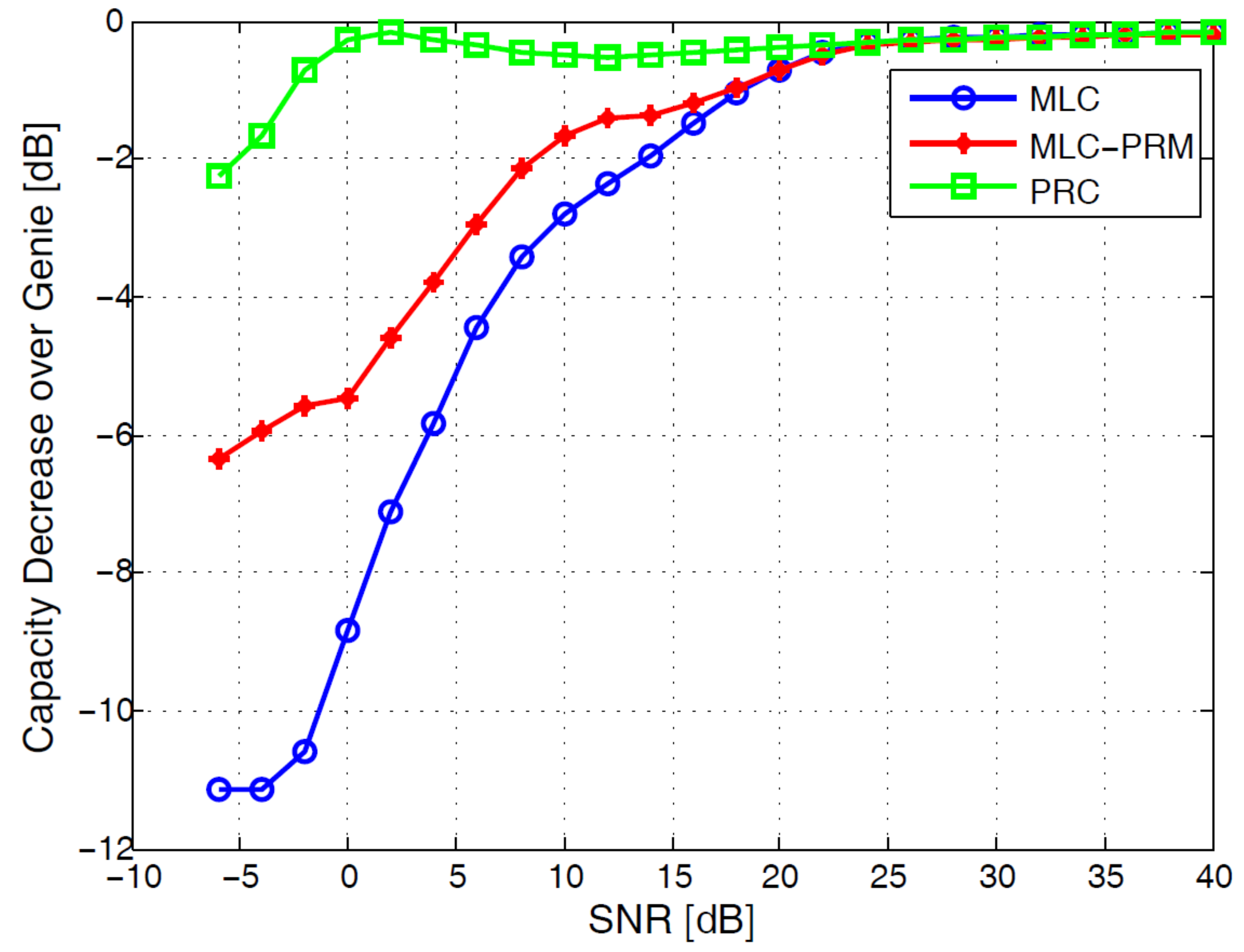}
		\caption{Capacity decrease of MLC, MLC-PRM, and PRC as compared to that of Genie in Case 3}
		\label{Fig:Capacity_Ex3}
	\end{figure}
	
	The capacity degradation due to incorrect signal classification, with respect to the Genie scheme with the assumption of ideal classification, is shown in Fig. \ref{Fig:Capacity_Ex3}. 
	When the SNR is lower than 20 dB, MLC-PRM and PRC obviously provide better user capacity than MLC, and in particular, PRC shows almost the same capacity as Genie.
	In the high SNR region, the classification rates of all schemes achieve almost the same capacity as the Genie scheme.
	
	\begin{figure} [t!]
		\centering
		\includegraphics[width=0.3\textwidth]{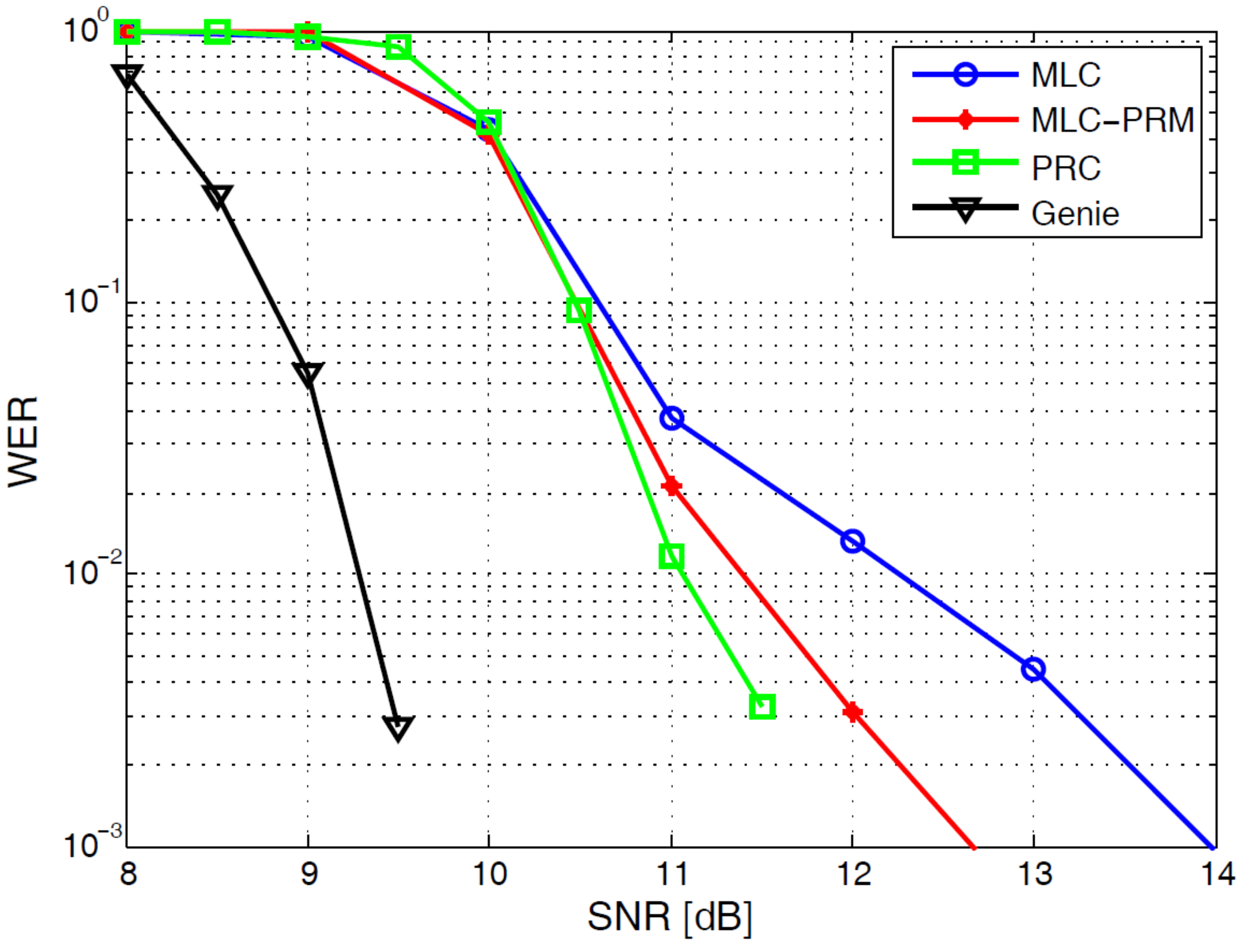}
		\caption{Comparison of WER performances in Case 1}
		\label{Fig:WER_Ex1}
	\end{figure}
	\begin{figure} [t!]
		\centering
		\includegraphics[width=0.3\textwidth]{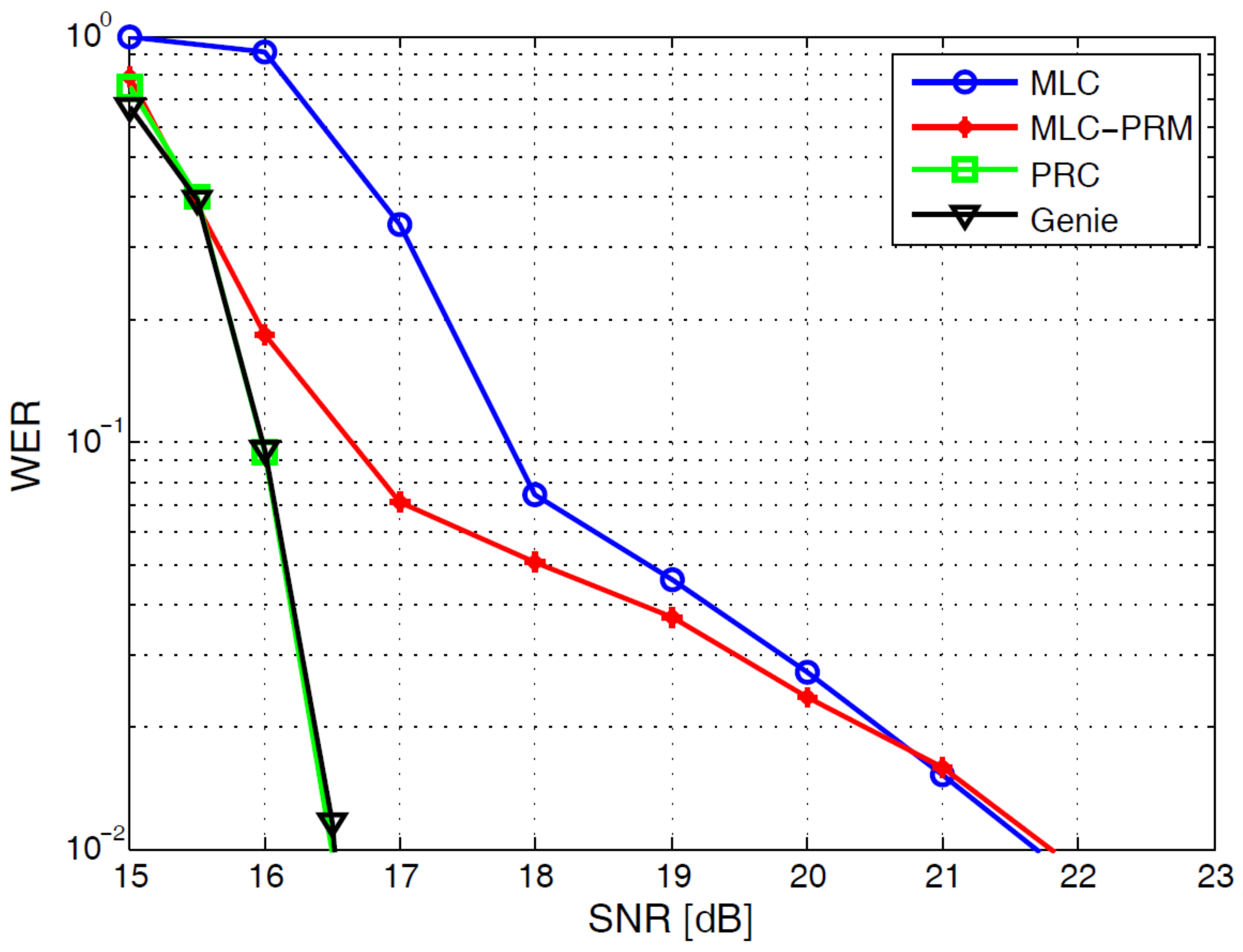}
		\caption{Comparison of WER performances in Case 2}
		\label{Fig:WER_Ex2}
	\end{figure}
	\begin{figure} [t!]
		\centering
		\includegraphics[width=0.3\textwidth]{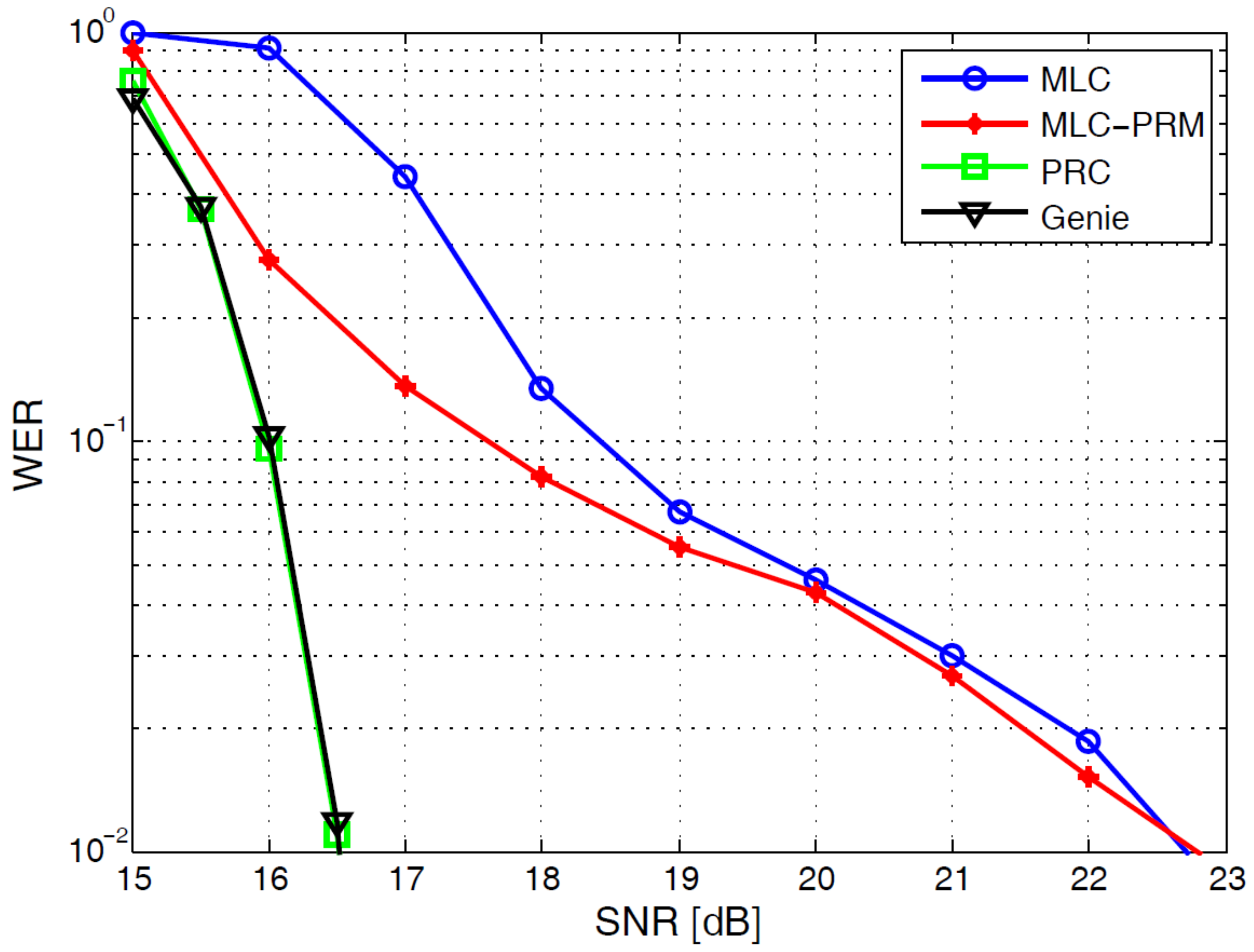}
		\caption{Comparison of WER performances in Case 3}
		\label{Fig:WER_Ex3}
	\end{figure}
	
	\subsection{Word Error Rates}
	
	The word error rate (WER) performances of the signal classification schemes are obtained to verify the practical usefulness of the proposed schemes.
	Figs. \ref{Fig:WER_Ex1}, \ref{Fig:WER_Ex2}, and \ref{Fig:WER_Ex3} show the WER performances of the comparison classification schemes in Cases 1, 2, and 3, respectively. 
	
	In Case 1, the WER performance of every signal classification scheme is significantly worse than that of Genie.
	In Fig. \ref{Fig:DetNOMANearErr_Ex1}, it can be seen that the classification error rates of none of the schemes are low enough to compete with the performance of Genie in the SNR region lower than 10 dB. 
	As the SNR increases above 10 dB, the near/far UT classification rates obtained by MLC-PRM and PRC eventually become superior to those obtained by MLC, and this tendency is reflected in Fig. \ref{Fig:WER_Ex1}.
	
	The WER graphs of Cases 2 and 3 shown in Figs. \ref{Fig:WER_Ex2} and \ref{Fig:WER_Ex3}, respectively, are quite similar.
	In these figures, the performance of PRC is nearly the same as that of the Genie scheme.
	It can be seen from Figs. \ref{Fig:DetNOMANearErr_Ex2} and \ref{Fig:DetNOMANearErr_Ex3} that PRC provides remarkably better classification error rates than MLC and MLC-PRM around 15 dB; In Figs \ref{Fig:WER_Ex2} and \ref{Fig:WER_Ex3}, PRC appears almost always to classify the NOMA signals correctly in the operating SNR region.
	
	MLC-PRM provides a 1 dB SNR gain at a WER of 0.1 as compared to MLC in Cases 2 and 3; however, the WER performances of PRC and Genie are still considerably better than those of MLC and MLC-PRM, even in the high SNR region. 
	These results are consistent with the MC error rates shown in Figs. \ref{Fig:DetNOMANearErr_Ex2} and \ref{Fig:DetNOMANearErr_Ex3}.
	In the high SNR region, the near/far UT classification error rates of all methods are sufficiently improved, and the effect of the MC error becomes dominant.
	Thus, the WER performance of PRC is much better than that of MLC and MLC-PRM in the high SNR region.
	However, because the MC error rates of MLC and MLC-PRM are identical, their WER performances converge.
	
	\section{Conclusions}
	
	In this paper, one of the key issues in NOMA systems, namely the blind signal classification problem of reducing high-layer signaling to provide information about the co-scheduled signal formats and to improve spectrum/resource efficiency in highly-mobile vehicular networks, was addressed. 
	We considered the classification steps of OMA/NOMA, near/far UT, modulation orders, and power ratios for NOMA UTs.
	In this study, the effects of each type of classification error in terms of SINR were quantified, and the capacity of the NOMA user was derived considering the signal classification errors.
	This paper proposed a phase-rotated modulation scheme and a pilot reuse-based signal classification that rotate data symbols and pilot symbols, respectively, and utilize the estimated phase to classify the received signal.
	The proposed schemes yield better performances in terms of classification error rate, capacity, and WER than conventional ML classification in various environment settings.
	Hence, the proposed schemes can be helpful in vehicular networks where only limited energy and spectrum/resource/time are available because of high mobility.

	\ifCLASSOPTIONcaptionsoff
	\newpage
	\fi

	%

	\begin{IEEEbiography}[{\includegraphics[width=1in,height=1.25in,clip,keepaspectratio]{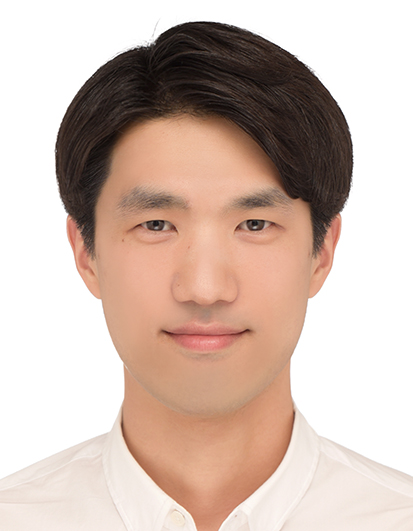}}]{Minseok Choi}
		is a postdoctoral researcher in the Department of Electrical and Computer Engineering at University of Southern California, Los Angeles, CA, USA, and also a postdoctoral associate in the School of Software at Chung-Ang University, Seoul, Korea.
		He received the B.S., M.S. and Ph.D. degrees in the School of Electrical Engineering from Korea Advanced Institute of Science and Technology (KAIST), Daejeon, Korea, in 2011, 2013, and 2018, respectively. 
		His research interests include wireless caching networks, stochastic network optimization, non-orthogonal multiple access and 5G networks.\end{IEEEbiography}
	\begin{IEEEbiography}[{\includegraphics[width=1in,height=1.25in,clip,keepaspectratio]{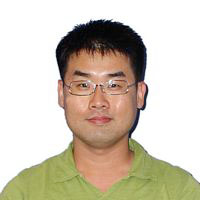}}]{Daejung Yoon}
		received the B.S. and M.S. degrees in electrical engineering from Kyungpook National University, South Korea, in 2003 and 2005, respectively. He received the M.S./Ph.D. degrees in wireless communication signal processing from the Electrical Engineering Department, the University of Minnesota, Minneapolis, Minnesota, USA, in 2011. In 2009, he joined Samsung Information Systems America, San Jose, California, where he worked on advanced signal processing and performance optimization. Since 2011, he has worked at Intel Cooperation, Santa Clara, California, and is extending his research field to include wireless communication system developments. At Intel, he is contributing to Intel LTE/LTE-A handset modem design and products. Since 2014, he has served as an Intel 3GPP standardization delegate in the 3GPP working group RAN4, and now he is a senior research specialist in Bell Labs Nokia, France working on 5G communication systems and 3GPP NR system design in release 16 and beyond. He has published multiple technical papers and patents on 5G NR cellular network systems and usecase studies’ as well as LTE/LTE-A evolution.\end{IEEEbiography}
	\begin{IEEEbiography}[{\includegraphics[width=1in,height=1.25in,clip,keepaspectratio]{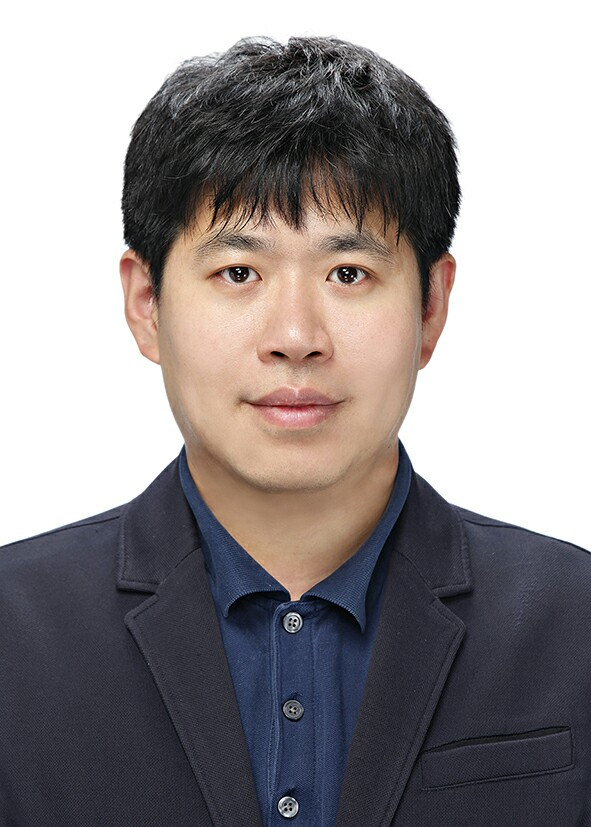}}]{Joongheon Kim}
		(M’06–SM’18) has been an assistant professor with Korea University, Seoul, Korea, since 2019. He received his B.S. (2004) and M.S. (2006) in computer science and engineering from Korea University, Seoul, Korea; and his Ph.D. (2014) in computer science from the University of Southern California (USC), Los Angeles, CA, USA. Before joing Korea University as a faculty member, he was with Chung-Ang University (Seoul, Korea, 2016--2019), Intel Corporation (Santa Clara, CA, USA, 2013–2016), InterDigital (San Diego, CA, USA, 2012), and LG Electronics (Seoul, Korea, 2006--2009). He is a senior member of the IEEE; and a member of IEEE Communications Society. He was awarded Annenberg Graduate Fellowship with his Ph.D. admission from USC (2009).
	\end{IEEEbiography}\vfill
	
	
	


\begin{thebibliography}{1}
		
		\bibitem{OMA_vs_NOMA:Wang}
		P. Wang, J. Xiao, and L. Ping, ``Comparison of Orthogonal and Nonorthogonal Approaches to Future Wireless Cellular Systems," {\it{IEEE Vehicular Technology Magazine}} vol. 1, no. 3, pp. 4--11, September 2006.
		
		\bibitem{TVT:1}
		Y. Lin, S. Wang, X. Bu, C. Xing, and J. An, ``NOMA-Based Calibration for Large-Scale Spaceborne Antenna Arrays," \textit{IEEE Transactions on Vehicular Technology}, vol. 67, no. 3, pp. 2231--2242, March 2018.
		
		\bibitem{TVT:2}
		Y. Gao, B. Xia, Y. Liu, Y. Yao, K. Xiao, and G. Lu, ``Analysis of the Dynamic Ordered Decoding for Uplink NOMA Systems With Imperfect CSI," \textit{IEEE Transactions on Vehicular Technology}, vol. 67, no. 7, pp. 6647--6651, July 2018.
		
		\bibitem{NOMA_basic:Docomo-Saito}
		Y. Saito, Y. Kishiyama, A. Benjebbour, T. Nakamura, A. Li, and K. Higuchi, ``Non-Orthogonal Multiple Access (NOMA)
		for Cellular Future Radio Access," in \textit{Proc. of IEEE Vehicular Technology Conference (VTC)}, Dresden, Germany, June 2013.
		
		\bibitem{NOMA_basic:Higuchi}
		K. Higuchi and Y. Kishiyama, ``Non-Orthogonal Access with Random Beamforming and Intra-Beam SIC for Cellular MIMO Downlink," in \textit{Proc. of IEEE Vehicular Technology Conference (VTC)}, Las Vegas, NV, USA, September 2013.
		
		\bibitem{NOMA_5G:Dai}
		L. Dai, B. Wang, Y. Yuan, S. Han, C. l. I, and Z. Wang, ``Non-Orthogonal Multiple Access for 5G: Solutions, Challenges, Opportunities, and Future Research Trends," \textit{IEEE Communications Magazine}, vol. 53, no. 9, pp. 74--81, September 2015.
		
		\bibitem{NOMA_5G:Ding}
		Z. Ding, Z. Yang, P. Fan, and H. V. Poor, ``On the Performance of Nonorthogonal Multiple Access in 5G Systems with
		Randomly Deployed Users," \textit{IEEE Signal Processing Letters}, vol. 21, no. 12, pp. 1501--1505, 
		December 2014.
		
		\bibitem{NOMA_5G:Islam}
		S. M. R. Islam, N. Avazov, O. A. Dobre, and K. S. Kwak, ``Power-Domain Non-Orthogonal Multiple Access (NOMA) in
		5G Systems: Potentials and Challenges", \textit{IEEE Communications Surveys \& Tutorials}, vol. 19, no. 2, pp. 721--742, June 2017.
		
		\bibitem{NOMA_VN:Di}
		B. Di, L. Song, Y. Li, and Z. Han, ``V2X Meets NOMA: Non-Orthogonal Multiple Access for 5G-Enabled Vehicular Networks," \textit{IEEE Wireless Communications}, vol. 24, no. 6, pp. 14--21, Dec. 2017.
		
		\bibitem{NOMA_VN:Qian}
		L. P. Qian, Y. Wu, H. Zhou, and X. Shen, ``Non-Orthogonal Multiple Access Vehicular Small Cell Networks: Architecture and Solution," \textit{IEEE Network}, vol. 31, no. 4, pp. 15--21, April 2017.
		
		\bibitem{MuST:3GPP}
		3GPP, ``Study on Downlink Multiuser Superposition Transmission for LTE," \textit{3GPP, TR 36.859}, January 2016.
		
		
		\bibitem{Book:Tse}
		D. Tse and P. Viswanath, \textit{Fundamentals of Wireless Communication}, Cambridge University Press, 2005.
		
		\bibitem{NOMA-MIMO:TWC2016Ding}
		Z. Ding, F. Adachi, and H. V. Poor, ``The application of MIMO to nonorthogonal multiple access,” \textit{IEEE Transactions on Wireless Communications}, vol. 15, no. 1, pp. 537-–552, Jan. 2016.
		
		\bibitem{NOMA-MIMO:TWC2016Ding2}
		Z. Ding, R. Schober, and H. V. Poor, ``A general MIMO framework
		for NOMA downlink and uplink transmission based on signal alignment,” \textit{IEEE Transactions on Wireless Communications}, vol. 15, no. 6, pp. 4438–-4454, Jun. 2016.
		
		\bibitem{NOMA:CL2015Ding}
		Z. Ding, M. Peng and H. V. Poor, ``Cooperative Non-Orthogonal Multiple Access in 5G Systems," \textit{IEEE Communications Letters}, vol. 19, no. 8, pp. 1462-1465, Aug. 2015.
		
		\bibitem{NOMA:TWC2018Choi}
		M. Choi, D. Han and J. Moon, ``Bi-Directional Cooperative NOMA Without Full CSIT," \textit{IEEE Transactions on Wireless Communications}, vol. 17, no. 11, pp. 7515-7527, Nov. 2018.
		
		\bibitem{IOTJ2019Liu2}
		M. Liu, J. Yang and G. Gui, ``DSF-NOMA: UAV-Assisted Emergency Communication Technology in a Heterogeneous Internet of Things," Early Access, \textit{IEEE Internet of Things Journal.}
			
		\bibitem{NOMA:TVT2018Gui}
		G. Gui, H. Huang, Y. Song and H. Sari, ``Deep Learning for an Effective Nonorthogonal Multiple Access Scheme," \textit{IEEE Transactions on Vehicular Technology}, vol. 67, no. 9, pp. 8440--8450, Sept. 2018.
		
		\bibitem{IOTJ2019Liu}
		M. Liu, T. Song and G. Gui, ``Deep Cognitive Perspective: Resource Allocation for NOMA based Heterogeneous IoT with Imperfect SIC," Early Access, \textit{IEEE Internet of Things Journal.}
			
		\bibitem{MC-ML-unknown1:Hameed}
		F. Hameed, O. A. Dobre, and D. C. Popescu, ``On the Likelihood-based Approach to Modulation Classification," \textit{IEEE Transactions on Wireless Communications}, vol. 8, no. 12, pp. 5884--5892, December 2009.
		
		\bibitem{MC-ML-unknown2:Erfan}
		E. Soltanmohammadi and M. Naraghi-Pour, ``Blind Modulation Classification over Fading Channels Using Expectation-Maximization," \textit{IEEE Communications Letters}, vol. 17, no. 9, pp. 1692--1695, September 2013.
		
		\bibitem{MC-ML:Wei}
		W. Wei and J. M. Mendel, ``Maximum-Likelihood Classification for Digital Amplitude-Phase Modulations," \textit{IEEE Transactions on Communications}, vol. 48, no. 2, pp. 189--193, February 2000.
		
		\bibitem{MC-feature:ICCSPA2013Hazza}
		A. Hazza, M. Shoaib, S. A. Alshebeili and A. Fahad, ``An overview of feature-based methods for digital modulation classification," \textit{2013 1st International Conference on Communications, Signal Processing, and their Applications (ICCSPA)}, Sharjah, 2013, pp. 1-6.
		
		\bibitem{MC-ML:TCCN2018Rajendran}
		S. Rajendran, W. Meert, D. Giustiniano, V. Lenders and S. Pollin, ``Deep Learning Models for Wireless Signal Classification With Distributed Low-Cost Spectrum Sensors," \textit{IEEE Transactions on Cognitive Communications and Networking}, vol. 4, no. 3, pp. 433--445, Sept. 2018.
		
		\bibitem{CNN-MC:DySPAN2017West}
		N. E. West and T. O'Shea, ``Deep architectures for modulation recognition," \textit{2017 IEEE International Symposium on Dynamic Spectrum Access Networks (DySPAN)}, Piscataway, NJ, 2017, pp. 1--6.
		
		\bibitem{CNN-MC:TVT2018Meng}
		F. Meng, P. Chen, L. Wu and X. Wang, ``Automatic Modulation Classification: A Deep Learning Enabled Approach," \textit{IEEE Transactions on Vehicular Technology}, vol. 67, no. 11, pp. 10760--10772, Nov. 2018.
		
		\bibitem{CNN-MC:TCCN2018Tian}
		J. Tian, Y. Pei, Y. Huang and Y. Liang, ``Modulation-Constrained Clustering Approach to Blind Modulation Classification for MIMO Systems," \textit{IEEE Transactions on Cognitive Communications and Networking}, vol. 4, no. 4, pp. 894--907, Dec. 2018.
		
		\bibitem{CommMag2013Araniti}
		G. Araniti, C. Campolo, M. Condoluci, A. Iera, and A. Molinaro, ``LTE for vehicular networking: A survey,” \textit{IEEE Communications Magazine}, vol. 51, no. 5, pp. 148–-157, May 2013.
		
		\bibitem{TVT2016Kim}
		J. Kim, S.-C. Kwon, and G. Choi, ``Performance of video streaming in infrastructure-to-vehicle telematic platforms with 60-GHz radiation and IEEE 802.11ad baseband,” \textit{IEEE Trans. Vehicular Technology}, vol. 65, no. 12, pp. 10111–-10115, Dec. 2016.
		
		\bibitem{TVT2017Majhi}
		S. Majhi, R. Gupta, W. Xiang and S. Glisic, ``Hierarchical Hypothesis and Feature-Based Blind Modulation Classification for Linearly Modulated Signals," \textit{IEEE Transactions on Vehicular Technology}, vol. 66, no. 12, pp. 11057--11069, Dec. 2017.
		
		\bibitem{ArXiv2019Choi}
		M. Choi and J. Kim, ``Blind Signal Classification Analysis and Impact on User Scheduling and Power Allocation in Nonorthogonal Multiple Access," available at: arxiv.org/abs/1902.02090
		
		\bibitem{VTC2015Yan}
		C. Yan, \textit{et. al.}, ``Receiver Design for Downlink Non-Orthogonal Multiple Access (NOMA)," in {\it{Proc. IEEE Vehicular Technology Conference (VTC),}} Glasgow, UK, May 2015.
		
		\bibitem{ldpc_11ad}
		A. K. Gupta, \textit{et. al.}, ``BER Performance of IEEE 802.11ad for Single Carrier and Multi Carrier," {\it{International Journal of Engineering Science and Technology}}, vol. 1, no. 4, pp. 2180--2187, May 2012.
		
	\end{thebibliography}
\end{document}